\documentclass[aps,12pt,showpacs,showkeys,nofootinbib]{revtex4-1}

\usepackage[english]{babel}
\usepackage{amsmath}
\usepackage{amssymb}
\usepackage{amsbsy}
\usepackage{amstext}
\usepackage{graphicx}
\usepackage{pst-node}
\usepackage{simplewick}
\usepackage{hyperref}
\usepackage{verbatim}
\usepackage{ulem}

\newcommand{\be}{\begin{eqnarray}}
\newcommand{\ee}{\end{eqnarray}}
\newcommand{\bdm}{\begin{displaymath}}
\newcommand{\edm}{\end{displaymath}}
\newcommand{\ds}{\displaystyle}
\newcommand{\ba}{\begin{array}}
\newcommand{\ea}{\end{array}}
\newcommand{\pa}[1]{\left(#1\right)}
\newcommand{\paq}[1]{\left[#1\right]}
\newcommand{\pag}[1]{\left\{#1\right\}}

\newcommand{\K}{{\bf k}}
\newcommand{\Q}{{\bf q}}
\newcommand{\pp}{{\bf p}}

\newcommand{\intkq}{\int_{\K\,,\Q}}

\begin{document}

\normalem
\title{Hereditary terms at next-to-leading order in two-body gravitational dynamics}

\author{Stefano Foffa$^{\rm 1}$ and Riccardo Sturani$^{\rm 2}$}

\affiliation{$(1)$ D\'epartement de Physique Th\'eorique and Center for Astroparticle Physics, Universit\'e de 
             Gen\`eve, CH-1211 Geneva, Switzerland\\
             $(2)$ International Institute of Physics, Universidade Federal do Rio Grande do Norte,
             Campus Universit\'ario, Lagoa Nova, Natal-RN 59078-970, Brazil}

\email{stefano.foffa@unige.ch, riccardo@iip.ufrn.br}

\begin{abstract}
In the context of the two-body problem in General Relativity, hereditary terms
in the long range gravitational field depend on the history rather than the
instantaneous state of the source at retarded time.
We compute the next-to leading effects of such hereditary terms, that comprise
tail and memory, on the two-body dynamics, within effective field 
theory methods, including both dissipative and conservative effects. 
The former confirm known results at 2.5 post-Newtonian
order with respect to the leading order in the luminosity function;
the conservative part is a new result and is an unavoidable
ingredient for a derivation of the conservative two-body dynamics at fifth
post-Newtonian order.
\end{abstract}



\maketitle

\section{Introduction}
\label{sec:intro}
The recent detections of Gravitational Waves (GWs)
(see \cite{LIGOScientific:2018mvr} for a summary of all confirmed detections
up to the time of writing),
beside marking the beginning of the new science named GW Astronomy, have triggered scientific
interest over all aspects of GW production and detection.

To maximize the efficiency of the search for signals from compact binary
coalescences, the output of GW detectors like
LIGO \cite{TheLIGOScientific:2014jea} and Virgo \cite{TheVirgo:2014hva} is
processed via matched-filtering \cite{Allen:2005fk}, which is particularly
sensitive to the phase of the GW signals.
This is a mixed blessing having the downside of faithful parameter
reconstruction depending on the availability of accurate model of signals, and the advantage
of offering a unique probe to the quantitative details of the highly non-linear
regime of General Relativity (GR).

One of the pillars to construct waveform templates for the LIGO/Virgo
data analysis pipeline has been the post-Newtonian (PN) approximation to GR,
see \cite{Blanchet:2013haa} for a review, which is a perturbative
method expanding the two-body dynamics around the Newtonian result, with
expansion parameter the relative velocity $v$, where $v^2\sim G_NM/r$ for Kepler
law
(with $G_N$ the Newton's constant, $M$ to total mass of the binary system, $r$
the binary constituent mutual distance, using natural units for the speed of light $c=1$), and $n$-PN corrections corresponding to terms of the order $G_N^{n-j+1}v^{2j}$,
with $0\leq j\leq n+1$.
To construct accurate waveform templates describing the entire coalescence,
including the merger of the two bodies, the PN-approximation must be completed
with non-perturbative results derived from numerical simulation, see
e.g.~\cite{Mroue:2013xna} for one of the most complete numerical waveform catalogs, which has brought to the
successful implementation of phenomenological model{\red s} \cite{Taracchini:2013rva,Pan:2013rra,Hannam:2013oca} merging information
from analytic and numerical relativity.
In particular the Effective Field Theory (EFT) approach to the PN approximation to GR adopted here,
pioneered in \cite{Goldberger:2004jt}, see \cite{Goldberger:2007hy,Foffa:2013qca,Porto:2016pyg,Levi:2018nxp} for reviews, has, among others, the undeniable
advantage of recasting the GR 2-body problem into the powerful language of
field theory scattering amplitudes, which has been developed and enriched of
deep theoretical insight over decades.

At present the dynamics is known in the conservative sector up to 4th PN order
\cite{Damour:2015isa,Damour:2016abl,Bernard:2015njp,Marchand:2017pir,Foffa:2012rn,Foffa:2016rgu,Foffa:2019rdf}, i.e. next-next-next-next-to-leading order
(N$^4$LO) for the spin-less terms, and up to 3.5PN and 4PN order \cite{Porto:2008tb,Porto:2008jj,Porto:2010tr,Porto:2010zg,Levi:2011eq,Porto:2012as,Levi:2015uxa,Levi:2015ixa,Levi:2014gsa,Levi:2016ofk} for terms including spin.
In the dissipative sector current knowledge of the luminosity function extends
to 3.5PN order for spin independent \cite{Blanchet:2001ax,Blanchet:2004ek} and
to 3.5PN for linear-in-spin  (N$^2$LO)\cite{Bohe:2013cla} terms
(and to 4PN order for tail and linear-in-spin terms (NLO)
\cite{Marsat:2013caa}, see below for tail definition),
up 3PN for terms quadratic in spins (NLO) \cite{Bohe:2015ana} and at leading order
(3.5PN) for spin cube effects \cite{Marsat:2014xea}.
The leading PN order for spin interaction to the $m$-th power for
both dissipative and conservative sector were computed in
\cite{Siemonsen:2017yux}, corresponding to $(1/2+m)$-PN order for $m$ odd and to
$(m)-$PN order for $m$ even.
In the spin-less case, the fifth PN order in the conservative sector (which is the main focus of this work) is qualitatively different from the lower ones, where finite
size effects cannot affect the dynamics
as per the \emph{effacement principle} \cite{Damour:1982wm}.
In the case of neutron star finite size effects are parametrized by tidal Love
numbers \cite{Damour:1982wm,Hinderer:2007mb}, whose first preliminary measure
has been enabled by the detection of GW170817
\cite{TheLIGOScientific:2017qsa,Abbott:2018wiz}, whereas for black holes it has
actually been shown
\cite{Binnington:2009bb,Damour:2009vw,Kol:2011vg,Gurlebeck:2015xpa}
that tidal deformation vanishes in the \emph{static} case, pushing its effect
to higher orders (see \cite{Pani:2015hfa,Pani:2015nua,Landry:2015zfa} for slowly
spinning black holes).

In the PN approach it is convenient to divide the problem of binary
dynamics in a \emph{near} and a \emph{far zone}: the former describes the
conservative dynamics around the sources at a distance $\sim r$ at which is
possible to resolve the individual constituents of the binary system,
the latter describes the dynamics from a much larger distance $> r/v=\lambda$
with $\lambda$ the GW wavelength, where the binary system can be described
by a single object endowed with multipoles.

An intriguing aspect of the post-Newtonian approximation is that near and far
zone are not disconnected:
while near zone results determine conservative dynamics only, far zone ones
give contribution to both conservative and dissipative dynamics, and their
contribution is necessary to obtain consistent results in the conservative
sector, in particular the unambiguous and systematic cancellation of spurious infra-red divergences in the
near zone \cite{Foffa:2019yfl}.

The contribution of far zone dynamics to conservative physics was first
observed in \cite{Blanchet:1987wq}, where the effect of GW emitted by the
binary system and scattered off the quasi-static curvature onto the same GW
source was (partially) computed and the name of \emph{tail} terms was coined to
indicate the ``back-scattering'' of GWs.
This process was understood as part of phenomena where the near-zone physics
depends on the full past history of the source, hence the name \emph{hereditary},
also coined in \cite{Blanchet:1987wq}, rather than just on the source state at retarded
time.
Hereditary terms affect the phasing of the gravitational waveform via the tail
effect \cite{Blanchet:1992br,Blanchet:1993ec}, see also \cite{Goldberger:2009qd,Porto:2010zg} for an
EFT derivation, but also via scattering off the curvature induced by GW
themselves, that is the \emph{memory} effect.
Memory effect causes a cumulative change in the waveform, that does not
vanish after the passage of the radiation,
originally derived in \cite{Christodoulou:1991cr}
and first derived in the binary system context in \cite{Blanchet:1992br}.\footnote{The presence of memory effects was noticed in linearized gravity
already in \cite{1974SvA....18...17Z}, where the passage of GWs
sourced by moving massing objects was identified to cause a \emph{permanent}
displacement between test particles, not fading away after the gravitational
perturbation as gone quiet. It was later quantified in
\cite{Thorne:1992sdb,Wiseman:1991ss} to leading order $1/r$ in linearized
gravity and found to relate the
difference in the gravitational radiative field at early and late times to the
source velocities at early and late times.}

The present work is adding another brick to the construction of a complete
precision gravity program that maximizes the physics output of GW detection, 
while at the same time providing further insight into intriguing theoretical
aspects of the general relativistic two-body dynamics.
More in detail, we present the original result of next-to-leading order
hereditary processes with no external radiation, as done in
\cite{Foffa:2011np,Galley:2015kus} at leading order,
from which it is possible to extract contributions to the conservative
dynamics and the luminosity function.
The leading tail contribution to the luminosity function, determined by the
imaginary part of the amplitude, is at 1.5PN order
with respect to the leading order quadrupole formula,
while the real part contributes at 4PN order to the conservative sector where
it has a divergent and a finite piece:
the former is regularized by properly adding similarly divergent terms from the
near zone dynamics \cite{Foffa:2019yfl}, see \cite{Manohar:2006nz} for a
general treatment of divergence cancellation in theories where momentum
integrals are separated in regions (like in the near-far zone case), dubbed
\emph{zero bin subtraction}; the latter turns from hereditary to instantaneous
when computed over circular orbits, contributing (beside rational terms) with
a logarithmic term to the energy of circular orbit, which has been
determined at 5PN order in \cite{LeTiec:2011ab} and in
\cite{Bini:2013rfa} from gravitational self-force computation \footnote{See also\cite{Barack:2010ny} for related work on 5PN logs.}.

The memory term gives no contribution to the
luminosity function, however it starts contributing to the conservative dynamics at 5PN
order, and its value, which is of the same order of the next-to-leading
tail effect, is computed in this work for the first time.
Note that differently from the tail effect, which is of hereditary type both
in the GW phase and in the conservative dynamics (for generic orbits),
the memory terms entering
the conservative dynamics are not hereditary, as they affect the gravitational
waveform $h$ via a non-local in time term which can be written as the
time integral of an instantaneous term, therefore giving an instantaneous
contribution to the energy which depends on the time derivative of $h$.

The plan of the paper is the following: in sec.~\ref{sec:method} we outline
the EFT method we use to perform the computation of the 5PN hereditary terms.
In sec.~\ref{sec:results} we detail the computation and present the result,
including the determination of several new, unpublished terms in the
conservative sector, while confirming previous findings in the dissipative one,
and we finally conclude in sec.~\ref{sec:discussion}.

\section{Method}
\label{sec:method}
On length scales larger than the orbital separation, the multipole moments of the binary system
are the relevant degrees of freedom when it comes to describe its interaction with the
gravitational field; the effective Lagrangian governing the dynamics of the system is \cite{Goldberger:2009qd}
\be
\ba{rcl}
\label{eq:mult}
\ds S_{mult}&=&-\ds\frac1{\Lambda} \left\{\int {\rm d}{\tau}\left[E
+\frac12\dot x^{\mu}L_{\alpha\beta}\omega^{\alpha\beta}_{\mu}\right.\right.\\
&&\ds\qquad\qquad-\frac12\sum_{n\geq 0}\pa{
c^{\cal I}_n{\cal I}^{\mu_1\dots\mu_n\alpha\beta}{\cal E}_{\alpha\beta;\mu_1\dots\mu_n}+
c^{\cal J}_n{\cal J}^{\mu_1\dots\mu_n\alpha\beta}{\cal B}_{\alpha\beta;\mu_1\dots\mu_n}}
\bigg]\Bigg{\}}\\
&\simeq&\ds\frac 1{\Lambda} \int {\rm d}t\paq{\frac 12 E h_{00}+\frac12 \epsilon_{ijk}L^i h_{0j,k}+\frac12 Q^{ij}{\cal E}_{ij}+\frac{1}{6} O^{ijk}{\cal E}_{ij,k}-\frac{2}3J^{ij}{\cal B}_{ij}+\dots}\,,
\ea
\ee

where $\Lambda$ is related to the $d$ dimensional gravitational constant $G_d$
by $\Lambda^2\equiv 1/(32\pi G_d)$.\footnote{Note that in eq.~(\ref{eq:mult})
    the angular momentum is defined as $\epsilon_{ijk}L^i=T^{0j}x^k-T^{0k}x^j$, i.e. with a minus sign with respect to the standard definition.}
In the first line of eq.~(\ref{eq:mult}) the mass $E$ and the
angular momentum $L_{\alpha\beta}$ coupling to the spin connection have been
singled out (neglecting total momentum since we assume to work in the
center-of-mass frame), and electric and magnetic multipoles of generic
order have been indicated with $\cal I$ and $\cal J$, respectively, and are coupled to the appropriate
curvature tensors.
In the second line we have expanded the metric around Minkowski and reported only terms needed up to next-to-leading order: the multipole series is an expansion
in powers of $v= r/\lambda$, being $r$ the size
of the source and $\lambda$ the length curvature scale of the gravitational
field coinciding with the GW-length, which is not independent on the size
of source and its internal velocity $v$.

We have also expanded at linear order in the gravitational perturbation
$h_{\mu\nu}$ around Minkowski spacetime, and made explicit space-time
decomposition (Latin indices running over spatial dimension only).

The gravitational field is to be evaluated at the center of mass of the system,
and the relevant electric and magnetic tensors components read
${\cal E}_{ij}\equiv R_{0i0j}\simeq\frac12\left(h_{00,ij}+\ddot{h}_{ij}-\dot{h}_{0i,j}-\dot{h}_{0j,i}\right)$ and ${\cal B}_{ij}\equiv\frac12\epsilon_{ikl}R_{0jkl}$, with
$R_{0jkl}\simeq\frac12\left(\dot{h}_{jk,l}-\dot{h}_{jl,k}+h_{0l,jk}-h_{0k,jl}\right)$,
being $R^\mu_{\nu\rho\sigma}$ the standard Riemann tensor.

The multipole moments, $E$, $\vec{L}$, $Q_{ij}$, $O^{ijk}$ are respectively
the energy, spin, mass quadrupole and octupole moments of the system, the last
two symmetric and traceless, and the current quadrupole moment is defined as
$J^{ij}\equiv-\frac12 \int (j^ix^j+j^ix^j)$
(with $x^i T^{0j}-x^j T^{0i}\equiv\epsilon^{ijk}j^k$, $T^{ij}$ denoting the energy
momentum tensor, implying that $J^{ij}$ is also traceless).

The explicit expression of the multipole moments in terms of the individual
constituent of the binary system will not be needed until next section, where we will derive the logarithmic energy shift for a binary system in circular orbit.
Such expressions can be determined by a \emph{matching} procedure, i.e. computing
the coupling of external gravitational field to the binary energy momentum
tensor \cite{Goldberger:2004jt,Foffa:2013qca}, so they include also the
contribution of the gravitational interaction at the orbital scale; the last
remark is relevant for $E$ and $Q_{ij}$ which need to be considered at
next-to-leading order.

Eq.~(\ref{eq:mult}) can be explicitly derived from the fundamental
coupling $T^{\mu\nu}h_{\mu\nu}$ via multiple derivations by parts and reiterated
use of the equations of motion;
here we retained only terms which do not vanish on the equations of motion.

\begin{figure}
\includegraphics[width=.9\linewidth]{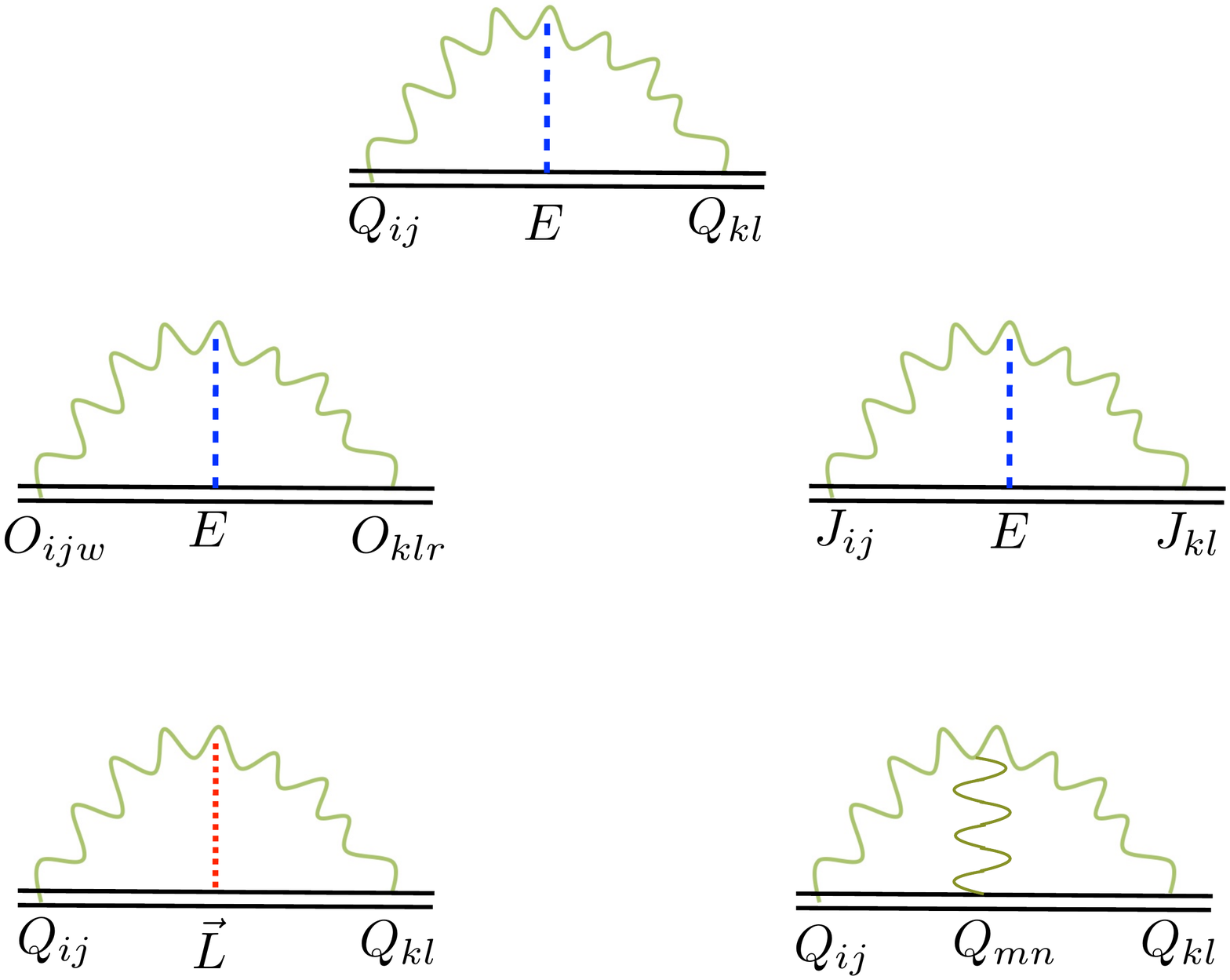}
\caption{Hereditary diagrams contributing from leading order
(top one), and from next-to leading order (the remaining four).
Wiggled lines represent on-shell gravitons (i.e. gravitational waves),
straight dashed and dotted ones stand for instantaneous propagators, i.e.
interaction with static fields. A relativistic correction to the instantaneous propagator of the top diagram has not been considered because it gives a physically irrelevant term (see discussion in the text).
The horizontal continuous black double line represent the external source given
by the binary system.}
\label{fig:hereditary}
\end{figure}


The present work reports the computation of a certain class of self-energy
diagrams, depicted in fig.~\ref{fig:hereditary}, representing self-energy
corrections due to the source interacting with the GWs produced by itself.
The imaginary part of these diagrams is related to
the power emission, while the real part gives their direct contribution to
the potential ruling the conservative dynamics.

We work in dimensional regularization and the real parts of some of the
hereditary diagrams present short-distance (UV) poles, which cancel against
long-scale (IR) spurious poles in the effective Lagrangian at the orbital scale
(near zone), according to the zero-bin prescription
\cite{Manohar:2006nz,Porto:2017shd,Porto:2017dgs} as explicitly shown at 4PN
order in \cite{Foffa:2019yfl}.
The finite terms remaining after such divergences cancellation depend
unambiguously on the finite contributions to the real parts of the hereditary
processes we show in the next section, and they are necessary to complete the
determination of the near zone dynamics.

The hereditary processes in fig.~\ref{fig:hereditary} involve a ``bulk''
three-field interaction that can be read from the (gauge-fixed) action for
gravity, which with our choice of the harmonic gauge reads 
\be
\label{eq:sEH}
{\mathcal S}_{EH+GF}&=&\frac1{16 \pi G_d}\int {\rm d}^{d+1}x\sqrt{-g}\left[R(g)-\frac 12\Gamma^\mu\Gamma_\mu\right]\,,
\ee
with $\Gamma^\mu\equiv g^{\nu\rho}\Gamma^\mu_{\nu\rho}$.
As in our previous works (see for instance \cite{Foffa:2019rdf} for details),
we find convenient to decompose the metric into a scalar $\phi$, a vector
$\vec{A}$ and a symmetric tensor $\sigma_{ij}$, which have the virtue of not
mixing with each other at quadratic order.
Expanding around Minkowski with the metric parametrization \cite{Kol:2007bc}
\allowdisplaybreaks
\be
\label{eq:ansatz}
g_{\mu\nu}=e^{2\phi/\Lambda}\pa{
\ba{cc}
\ds -1 &\ds \frac{A_j}\Lambda\\\ds
\frac{A_i}\Lambda &\ds e^{-c_d\phi/\Lambda}\pa{\delta_{ij}+\frac{\sigma_{ij}}\Lambda}-
\frac{A_iA_j}{\Lambda^2}
\ea}\,,
\ee
with $c_d\equiv 2(d-1)/(d-2)$, one obtains the following action, truncated to cubic order
\renewcommand{\arraystretch}{1.4}
\be
\label{eq:sEH_KK}
\ba{rcl}
\ds {\mathcal S}_{EH+GF} &\supset &\ds \int {\rm d}^{d+1}x\sqrt{-\gamma}
\left\{\frac{1}{4}\left[(\vec{\nabla}\sigma)^2-2(\vec{\nabla}\sigma_{ij})^2-\left(\dot{\sigma}^2-2(\dot{\sigma}_{ij})^2\right){\rm e}^{\frac{-c_d \phi}{\Lambda}}\right]- c_d \left[(\vec{\nabla}\phi)^2-\dot{\phi}^2 {\rm e}^{-\frac{c_d\phi}{\Lambda}}\right]\right.\\
&&\ds
+\left[\frac{F_{ij}^2}{2}+\left(\vec{\nabla}\!\!\cdot\!\!\vec{A}\right)^2 -\dot{\vec{A}}^2 {\rm e}^{-\frac{c_d\phi}{\Lambda}} \right]
{\rm e}^{\frac{c_d \phi}{\Lambda}}+\frac 2\Lambda\paq{\pa{F_{ij}A^i\dot{A^j}+\vec{A}\!\!\cdot\!\!\dot{\vec{A}}(\vec{\nabla}\!\!\cdot\!\!\vec{A})}
-c_d\dot{\phi}\vec{A}\!\!\cdot\!\!\vec{\nabla}\phi}\\
&&\ds
+2 c_d \left(\dot{\phi}\vec{\nabla}\!\!\cdot\!\!\vec{A}-\dot{\vec{A}}\!\!\cdot\!\!\vec{\nabla}\phi\right)
+\frac{\dot{\sigma}_{ij}}{\Lambda}\left(-\delta^{ij}A_l\hat{\Gamma}^l_{kk}+ 2A_k\hat{\Gamma}^k_{ij}-2A^i\hat{\Gamma}^j_{kk}\right)\\
&&\ds
-\left.\frac{1}{\Lambda}\left(\frac{\sigma}{2}\delta^{ij}-\sigma^{ij}\right)
\left({\sigma_{ik}}^{,l}{\sigma_{jl}}^{,k}-{\sigma_{ik}}^{,k}{\sigma_{jl}}^{,l}+\sigma_{,i}{\sigma_{jk}}^{,k}-\sigma_{ik,j}\sigma^{,k}
\right)\right\}\,,
\ea
\ee
\renewcommand{\arraystretch}{1.}
where $F_{ij}\equiv A_{j,i}- A_{i,j}$, and $\hat{\Gamma}^i_{jk}$ is the connection of the purely spatial $d$-dimensional metric $\gamma_{ij}\equiv\delta_{ij}+\sigma_{ij}/\Lambda$,
which is also used above to raise and contract spatial indices.
All spatial derivatives are understood as simple (not covariant) derivatives
and when ambiguities might raise gradients are always meant to act on
contravariant fields, e.g.
$\vec{\nabla}\!\!\cdot\!\!\vec{A}\equiv\gamma^{ij}A_{i,j}$
and $F_{ij}^2\equiv\gamma^{ik}\gamma^{jl}F_{ij}F_{kl}$.

In general, the amplitude for a generic hereditary process $A_{her}$ has the
following structure:
\be
\label{eq:her_gen}
A_{her}=\intkq \frac{{\rm d}q_0}{2\pi}\frac{{\rm d}k_0}{2\pi}M^{(1)}_{i_1\ldots i_l}(q_0)
M^{(2)}_{j_1\ldots j_m}(k_0)M^{(3)}_{k_1\ldots i_n}(-k_0-q_0)
\frac{P^{i_1\ldots i_lj_1\ldots j_mk_1\ldots k_n}(\K,\Q,k_0,q_0)}{\cal D}\,,
\ee
with $M^{(i)}_{i_1\ldots i_n}$ being the (Fourier-transformed) generic multipole
moment and we introduced the notation
$\int_\pp\equiv \int \frac{{\rm d}^dp}{(2\pi)^d}$, while the (inverse of the) factor ${\cal D}\equiv({\bf k}^2-k_0^2)({\bf q}^2-q_0^2)[({\bf k}+{\bf q})^2-(k_0+q_0)^2]$
collects the product of the scalar parts of the three propagators involved.
Also some more elementary integrals involving only two factors in the
denominator are involved in amplitude computations, and they are reported in
app.~\ref{app:integralacci}.

In the case of tail integrals one of the sources is actually conserved
(all but the bottom right diagram in fig.~\ref{fig:hereditary}) and substituting 
$M^{(1)}_{i_1\ldots i_n}(q_0)\to 2\pi \delta(q_0)\tilde M_{i_i\ldots i_n}$\footnote{Such substitution is identically true for the diagrams of the second line of figure \ref{fig:hereditary}, for which $E\simeq M$ so $\dot{M}=0$ identically, while for the upper diagram and for the lower left one this is true only modulo terms which vanish on the equations of motion, which are neglected here because they do not give contribution to gauge-invariant quantities (as they can be removed by an unphysical coordinate shift).} the amplitude
simplifies as one propagator become instantaneous:
\be
\label{eq:tail_gen}
A_{tail}=\intkq \frac{{\rm d}k_0}{2\pi}\tilde M_{i_1\ldots i_l}
M^{(2)}_{j_1\ldots j_m}(k_0)M^{(3)}_{k_1\ldots i_n}(-k_0)
\frac{P^{i_1\ldots i_lj_1\ldots j_mk_1\ldots k_n}(\K,\Q,k_0,0)}{{\cal D}|_{q_0=0}}\,.
\ee

In this case we can use Feynman boundary conditions for the propagators,
which give the (time-symmetric) real
Lagrangian contributing to the near zone
conservative dynamics, whereas the imaginary part returns the averaged
probability loss (related to the energy loss).
Had we been interested in computing back-reaction force or instantaneous
radiation field, we should have used in-in correlators as explained in detail
in \cite{Galley:2009px}.
On the other hand, when all three sources are dynamical, we will resort to retarded boundary conditions, see \cite{FS21} for detailed explanations.

Given these premises, all the integrals can be reduced in terms of the following
master integral
\be
\label{eq:master_int}
I_m (k_0,q_0) \equiv \intkq \frac 1{\cal D}\,,
\ee
plus other more elementary ones. $I_m$ is UV-divergent and has been extensively studied in particle physics, see
e.g. \cite{Davydychev:1992mt},
as it is the master integral of the two-loop vacuum diagram, also relevant
for two-loop self-energy diagrams in gauge theories.
We note however that only the specific case $I_m(k_0,0)$ (still UV-divergent) appears in our final results, namely in the first three diagrams of figure \ref{fig:hereditary};
this happens when the diagram is UV-divergent but one of the three sources is conserved, as in
eq.~(\ref{eq:tail_gen}). On the other hand
$I_m(k_0,q_0)$ appears only in some intermediate steps of the bottom right diagram and cancels in the final answer because such process is UV-finite.

\section{Results}
\label{sec:results}

\subsection{General properties of tails}
\allowdisplaybreaks
We start by reporting the result of the first amplitude (top of
fig.~\ref{fig:hereditary}), which appears at leading order
(1.5PN for power emission, and 4PN for the conservative part),
and has been already
considered within the EFT approach in \cite{Foffa:2011np, Galley:2015kus}.
At linear order the electric part of the Riemann tensor reads
\be
\label{eq:riem_el}
R_{0i0j}=\frac 12\ddot\sigma_{ij}-\frac 12\dot A_{i,j}-\frac 12 \dot A_{j,i}
-\phi_{,ij}-\frac{\delta_{ij}}{d-2}\ddot\phi+O(h^2)\,.
\ee
The leading tail amplitude is represented by the top diagram in
fig.~\ref{fig:hereditary} and its calculation is detailed here below
(omitting the propagator pole displacement in the complex plane, which is
understood to follow Feynman prescription)
and decomposed for polarizations: the only gravity
polarization coupling to the conserved energy is $\phi$, thus one has
six possibilities in terms of different polarizations for the
three-point vertex.

After neglecting terms proportional to quadrupole traces
the leading tail amplitude reads:
\be
\label{eq:tailQQ}
\ba{rll}
\ds iS_{eff\,4PN}^{E Q^2}=
-i 64 \pi^2 G^2_d E&
\ds\int_{-\infty}^\infty \frac{{\rm d}k_0}{2\pi}Q^{ij}(k_0)Q^{kl}(-k_0)\ds\intkq\,
\frac 1{\K^2-k_0^2}\frac 1{(\K+\Q)^2-k_0^2}\frac 1{\Q^2}\times &\\
&\ds\left\{
-\frac 18k_0^6\pa{\delta_{ik}\delta_{jl}+\delta_{il}\delta_{jk}}\right.&
\{\phi\sigma^2\}\\
&\ds -k_0^4\pa{k_i q_k\delta_{jl}}&\{\phi A \sigma\}\\
&\ds +\frac {k_0^2}{c_d}q_i k_j k_k k_l &\{\phi^2\sigma\}\\
&\ds +\frac 12k_0^2\paq{k_ik_jq_kq_l-q_ik_jk_kq_l+
\delta_{ik}k_j(k+q)_l\K\cdot(\K+\Q)}&\{\phi A^2\}\\
&\ds -\frac 1{c_d}k_0^2 q_k (k+q)_l k_i k_j &\{\phi^2 A\}\\
&\ds -\frac 1{2c_d}k_0^2 k_ik_j(k+q)_k(k+q)_l\Big\}&\{\phi^3\}\,,
\ea
\ee
giving the result
\be
\label{eq:tailQQ_res}
S_{eff\,4PN}^{E Q^2}&=&\ds -\frac{G_N^2E}5
\int_{-\infty}^\infty\frac{{\rm d}k_0}{2\pi}\,k_0^6
\ds\paq{\frac 1\epsilon-\frac{41}{30}-i\pi+\log\left(\frac{k_0^2{\rm e}^\gamma}{\pi\mu^2}\right)+O(\epsilon)}Q_{ij}(k_0)Q^{ij}(-k_0)\,,
\ee
where $\epsilon\equiv d-3$  and $\mu$ is the dimensional constant introduced in 
dimensional regularization to relate standard 4-dimensional Newton constant
$G_N$ to the $d-$dimensional gravitational coupling $G_d\equiv\mu^{-\epsilon}G_N$.
The factor $-\frac{41}{30}$, first derived in \cite{Foffa:2011np}, enables to
unambiguously determine the regularized near zone Lagrangian at 4PN as predicted in \cite{Porto:2017dgs} and explicitly done in \cite{Marchand:2017pir,Foffa:2019yfl}.

From the imaginary part\footnote{Differently from \cite{Galley:2015kus}, which presents the result in terms of the $(+,-)$ variables if the in-in formalism, the imaginary part of (\ref{eq:tailQQ_res}) does not present the term sgn($k_0$).} of eq.~(\ref{eq:tailQQ_res}) the power loss $P_{tail}$
can be derived by multiplying the integrand by $k_0$ and averaging over time.
The leading order power loss is the quadrupole formula $P_{QQ}$ which can be
obtained by the imaginary part of the following diagram
\be
\label{eq:PQQ}
\ba{rcl}
S^{Q^2}_{eff2.5PN}&=&\ds \int_{-\infty}^\infty\frac{{\rm d}k_0}{2\pi}\int_\K\Big(
\begin{minipage}{2.5cm}
\begin{center}
\includegraphics[width=\linewidth]{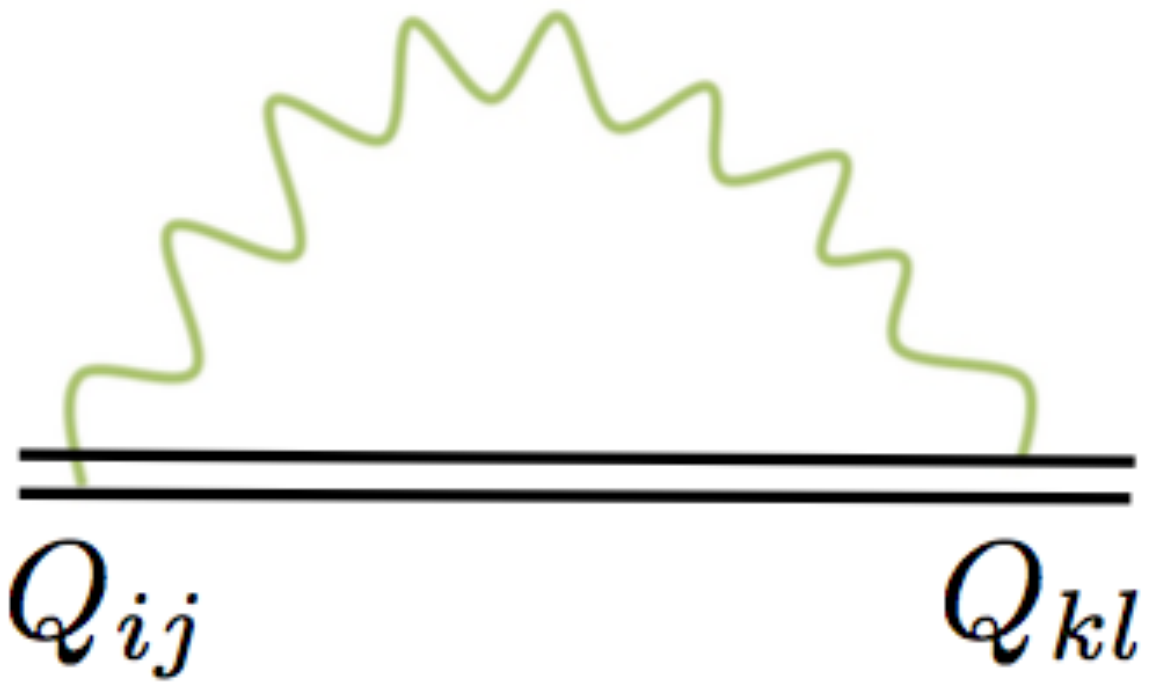}
\end{center}
\end{minipage}\Big)\\
&=&\ds -\pi G_N\int_{-\infty}^{\infty}\frac{{\rm d}k_0}{2\pi}
\int_\K\frac{Q_{ij}(k_0)Q_{kl}(-k_0)}{\K^2-k_0^2}
\paq{-k_0^4\delta^{ik}\delta^{jl}+2 k_0^2 \delta^{ik}k^{j}k^{l}
-\frac 12k^ik^jk^kk^l}\\
&=&\ds i\frac {G_N}{10}\int_{-\infty}^\infty\frac{{\rm d}k_0}{2\pi}|k_0|k_0^4 Q_{ij}(k_0)
Q^{ij}(-k_0)\,,
\ea
\ee

where the three terms in between round brackets,
$A_\sigma^{ijkl}\equiv -k_0^4\delta^{ik}\delta^{jl}$,
$A_A^{ijkl}\equiv 2k_0^2\delta^{ik}k^j k^l$,
$A_\phi^{ijkl}\equiv -k^ik^jk^kk^l/2$ (apart from a common normalization)
are the contributions respectively
from the $\sigma,A,\phi$ polarizations.
The result of $\K$ integration, see app.~\ref{app:integralacci}, has vanishing
real part and receives a finite imaginary part from the region of integration
where $|\K|\sim|k_0|$.

In the tail diagram, the double integration over space
momenta $\K,\Q$ on purely heuristic arguments leads to an amplitude result
$\propto G_d^2(-k_0^2)^{d-3}$
from which one can infer that the presence of an imaginary part is invariably
linked to a divergence:
$$\frac{G^2_d}\epsilon(-k_0^2-i0^+)^\epsilon=
G_N^2\pa{\frac 1\epsilon+\log(k_0^2/\mu^2)-i\pi+O(\epsilon)}\,,$$
implying that pole residual not only fixes the logarithmic term but also the
imaginary one.

Focusing on the divergent part of the tail amplitude,
it receives contributions only from processes involving the same graviton
polarizations attaching to the two radiative sources, as they are the ones diverging when $\Q\to 0$ (see app.~\ref{app:integralacci} for explicit
computations), and it can be written as
\be
\label{eq:pole_tail}
\ba{rl}
S_{pole}^{tail}=-64\pi^2G_N^2E&\ds\int_{-\infty}^\infty\frac{{\rm d}k_0}{2\pi}
\intkq\frac{k_0^2 Q_{ij}(k_0) Q_{kl}(-k_0)}
{\Q^2\left[(\K+\Q)^2-k_0^2\right]\pa{\K^2-k_0^2}}\\
&\ds\times\frac14\paq{-k_0^4 \delta^{ik}\delta^{jl}+2\K^2\delta^{ik}k^jk^l
-\frac 12k^ik^kk^jk^l}\,.
\ea
\ee
Note that the terms in square brackets in (\ref{eq:PQQ}) and
(\ref{eq:pole_tail}) are the same, apart from the substitution $\K^2\to k_0^2$ 
in $A_A^{ijkl}$, and the $\Q$ integration of the two propagators involving $\Q$
factorizes from the rest of the amplitude
with result $(-k_0^2)^{d/2-2}f(\K^2/k_0^2)$ for some function
$f$, which is the \emph{same for all multipoles} as they do not depend on $\Q$, thus giving:
\be
\label{eq:tailf}
\ba{rl}
\ds S_{pole}^{tail}=-64\pi^2G_N^2E\int_{-\infty}^\infty\frac{{\rm d}k_0}{2\pi}
(-k_0^2)^{d/2-1}&\ds\int_\K f\pa{\frac{\K^2}{k_0^2}}
\frac{Q_{ij}(k_0)Q_{kl}(-k_0)}{\K^2-k_0^2}\\
&\ds\times\paq{A_\sigma^{ijkl}+\frac{\K^2}{k_0^2}A_A^{ijkl}+A_\phi^{ijkl}}\,.
\ea
\ee
Now observing that the imaginary part must originate from the $\K^2=k_0^2$
region of integration, it follows that $f(\K^2/k_0^2)$ in (\ref{eq:tailf})
reduces to a UV divergent factor common to all multipoles (see appendix
\ref{app:integralacci} for details), hence it can be fixed by its value for
the quadrupole case.

In particular assuming that the
fundamental source is composed by a binary system with reduced mass $\eta M$
($\eta$ is the \emph{symmetric mass ratio}) on a circular orbit of radius $r$
and orbital angular velocity $\omega$, so that the quadrupole
component
$Q_{xx}(k_0)=\eta M r^2\pi/2\pa{\delta(k_0-2\omega)+\delta(k_0+2\omega)}$
and using $\delta(0)=T/(2\pi)$, the leading order Kepler law $G_NM/r=(r\omega)^2$
and the integral in eq.~(\ref{eq:int0}), one can derive the tail corrected
quadrupole formula
\be
\label{eq:QQ+tail}
\ba{rcl}
P_{QQ+tail}&=&\ds\frac{\eta^2G_NM^2r^4}{5} Im\paq{\int_0^\infty {\rm d}k_0
\pa{-k_0^2}^{1/2}k_0^4\delta(k_0-2\omega)\pa{1+2\pi G_N E k_0}}\\
&=&\ds\frac{32\eta^2}{5G_N}x^5(1+4\pi x^{3/2})\,,
\ea
\ee

where in the final step we have introduced the standard post-Newtonian
expansion parameter $x\equiv(G_N M\omega)^{2/3}$.
In the previous formula (\ref{eq:QQ+tail}) the factor $(1+2\pi G_NEk_0)$ is the
\emph{universal} leading tail correction for all multipoles\footnote{This includes also magnetic multipoles, for which a similar calculation can also be performed.}, universality
already noticed in \cite{Blanchet:1995fr} and in \cite{Goldberger:2009qd}.

Beside the imaginary term, clearly also the logarithmic term is fixed by the
tail divergent piece, which we now understand to be just $2 G_N E k_0$ times the
leading order imaginary part for all multipoles, hence
we have proven how to compute the far zone logarithmic
contribution to the conservative binary dynamics at $n$-PN order by the
result for the flux at $(n-4)$-PN order, as suggested  in \cite{Damour:2015isa}.
As a consequence of this universality, one can write down the action for \emph{all} the non-local simple tails
(we are not considering composite effects like tails of tails here), where the coefficient of each term is given by the coefficient of the corresponding
non-tail process in the power emission formula  $P=\frac15 \dddot{Q}_{ij}^2+\frac1{189}\ddddot{O}_{ijk}^2+\frac{16}{45} \dddot{J}_{ij}^2\dots$  \cite{Thorne:1980ru}:
\be
\label{eq:tail_univ}
\ba{rcl}
\ds{\cal S}^{tail}_{log}&=&\ds -G_N^2E\int_{-\infty}^\infty
\frac{{\rm d}k_0}{2\pi}\log\pa{\frac{k_0^2}{\mu^2}}\sum_{n\geq2}\,k_0^{2(n+1)}
c^n_{({\cal I},{\cal J})}({\cal I},{\cal J})^{\alpha\beta\mu_1\dots\mu_{n-2}}(k_0)({\cal I},
{\cal J})_{\alpha\beta\mu_1\dots\mu_{n-2}}(-k_0)\\
&=&\ds -2G_N^2 E \int_{-\infty}^{\infty} {\rm d}t
\sum_{n\geq2}(-1)^nc^n_{({\cal I},{\cal J})}({\cal I},{\cal J})^{\alpha\beta\mu_1\dots\mu_{n-2}}(t)
\int_0^{\infty}{\rm d}\tau  ({\cal I},{\cal J})_{\alpha\beta\mu_1\dots\mu_{n-2}}^{(2n+3)}(t-\tau) \log{(\mu\tau)}\,
\ea
\ee
with
\be
\ba{rcl}
\ds c^n_{\cal I}&=&\ds\frac{(n+1)(n+2)}{n (n-1) n! (2n+1)!!}\,,\\
\ds c^n_{\cal J}&=&\ds\frac{4 n (n+2)}{(n-1)(n+1)! (2n+1)!!}\,.
\ea
\ee

\subsection{Next-to-leading order hereditary terms}

In this subsection we compute amplitudes giving hereditary effects at NLO,
i.e. which start contributing to the power emission at 2.5PN order (unless they are
vanishing) and at 5PN for the conservative part.

\subsubsection{Octupole tail}
Here we present the computation of the tail-octupole amplitude (upper right
diagram of fig.~\ref{fig:hereditary})
\be
\label{eq:O_Kspace}
\ba{rlll}
\ds iS_{eff\, 5PN}^{M O^2}&\ds =
-i\frac{64}{9} \pi^2 G_d^2E&
\ds\int \frac{{\rm d}k_0}{2\pi}O^{ijw}(k_0)\,O^{klr}(-k_0)
\intkq\,\frac{k_w}{\K^2-k_0^2}\frac{(k+q)_r}{(\K+\Q)^2-k_0^2}&\ds\!\!\frac 1{\Q^2}\times\\
&&\ds\left\{
-\frac 18k_0^6\pa{\delta_{ik}\delta_{jl}+\delta_{il}\delta_{jk}}\right.&
\pag{\phi\sigma^2}\\
&&\ds -k_0^4\pa{k_j q_k\delta_{il}}&\pag{\phi A \sigma}\\
&&\ds +\frac{k_0^2}{c_d}q_l k_kk_i k_j&\pag{ \phi^2\sigma}\\
&&\ds \frac 12k_0^2\paq{k_i q_k-k_k q_i+
\delta_{ik}\K\cdot(\K+\Q)}k_j(k+q)_l&\pag{ \phi A^2}\\
&&\ds -\frac 1{c_d}k_0^2 k_i k_j q_k (k+q)_l&\ds\pag{\phi^2 A}\\
&&\ds-\left.\frac 1{2c_d}k_0^2k_ik_j(k+q)_k(k+q)_l\right\}&\pag{\phi^3}\,,
\ea
\ee
which adds to
\be
\label{eq:radReacResO}
S_{eff\,5PN}^{M O^2}&=&\ds -\frac{G_N^2M}{189}\int_{-\infty}^\infty
\frac{{\rm d}k_0}{2\pi}\,k_0^8\ds\paq{\frac 1\epsilon -\frac{82}{35}-i\pi+
\log\pa{\frac{k_0^2{\rm e}^\gamma}{\pi\mu^2}}+O(\epsilon)}
O_{ijw}(k_0)O^{ijw}(-k_0)\,.\nonumber\\
\ee
Also in this case, the imaginary part respects the universality of tail
terms described in the previous subsection, while the new finite real
coefficient $-\frac{82}{35}$, analogous to the $-\frac{41}{30}$ of the
quadrupole tail, is the finite correction to the near zone
conservative binary dynamics originally derived in this paper.
Note also that since this diagram is considered here at leading order, we are
entitled to trade $E$ with the total rest mass $M$ in the result.

\subsubsection{Magnetic quadrupole tail}

The second diagram in the second line in fig.~\ref{fig:hereditary} represents
the current quadrupole tail, which couples to the magnetic component of the
Riemann tensor:
\be
R_{0jkl}=\frac 12\left(\dot{\sigma}_{jk,l}-\dot\sigma_{jl,k}\right)+\frac 12\left(A_{l,kj}- A_{k,lj}\right)
+\frac1{d-2}\left(\dot{\phi}_{,k}\delta_{jl}-\dot{\phi}_{,l}\delta_{jk}\right) +O(h^2)\,.
\ee
A simplification happens here, as all contributions with a $\phi$ polarization
emitted by the magnetic quadrupole vanish, giving the result
\be
\label{eq:MQ_Kspace}
\ba{rll}
\ds iS^{M J^2}_{eff\,5PN}=&\ds-i\frac{1024}9 \pi^2G_d^2 E
\int \frac{{\rm d}k_0}{2\pi}J^{ij}(k_0)\,J^{kl}(-k_0)\ds\intkq\,
\frac{1}{\K^2-k_0^2}\frac{1}{(\K+\Q)^2-k_0^2}&\ds\!\!\frac 1{\Q^2}\times\\
&\ds\left\{
-\frac 18k_0^4\paq{\delta_{jl}\pa{\delta_{ik}\K\cdot(\K+\Q)-k_i(k+q)_k
+\epsilon_{jbk}\epsilon_{lai}k^b(k+q)^a}}\right.&\pag{\phi\sigma^2}\\
&\ds + \frac 14 k_0^2 k^j k^b (k+q)^a
\paq{q^c \epsilon_{ilb} \epsilon_{kac} + q^l\pa{\delta_{bk}\delta_{ai}-\delta_{ab}\delta_{ik}}}&\ds\pag{\phi A \sigma}\\
&\ds+\frac 18\epsilon_{i\alpha\beta}\epsilon_{k\gamma\delta}k_j k_\beta(k+q)_l(k+q)_\delta
\paq{\delta_{\alpha\gamma}\K\cdot(\K+\Q)-k_\gamma q_\alpha+k_\alpha q_\gamma}&
\pag{\phi A^2}\\
&\ds+0\Big\}\qquad\qquad\pag{\phi^2\sigma}\quad\{\phi^2 A\}\quad\{\phi^3\}\,.&
\ea
\ee
The result of this amplitude is
\be
\label{eq:radReacResJ}
S_{eff\,5PN}^{MJ^2}&=&\ds -\frac{16}{45} G_N^2M\int_{-\infty}^\infty
\frac{{\rm d}k_0}{2\pi}\,k_0^6\ds\pa{\frac 1\epsilon-\frac{127}{60}-i\pi
+\log\pa{\frac{k_0^2{\rm e}^\gamma}{\pi\mu^2}}+O(\epsilon)}
J_{ij}(k_0)J^{ij}(-k_0)\,,\nonumber\\
\ee
and we find again the expected coefficients of pole, logarithmic term and imaginary part, as well as a
second finite correction to the conservative dynamics, identified
by the rational number $-\frac{127}{60}$.

\subsubsection{Logarithmic contribution to energy of circular orbit}

The finite, instantaneous corrections to the conservative dynamics computed
above ($\frac{82}{6615}\ddddot O_{ijk}^2+\frac{508}{675}\dddot J_{ij}^2$)
affect the conservative dynamics of the binary system, once expressed
the multipoles in terms of individual binary constituent dynamical variables.

The hereditary logarithmic terms from the tail processes also become
instantaneous when the generic multipoles are specialized to a binary
system in \emph{circular} orbit and then give finite, logarithmic
contributions to the energy.

We expect that such logarithmic terms do not receive contributions from the near zone,  as it happens at 4PN order,
meaning that tail logarithms embody all of the logarithmic contribution to the energy of
circular orbit at 5PN. Using the 1PN corrected expression of the quadrupole
moment (see \cite{Goldberger:2009qd}) and the leading PN order of octupole and
magnetic moments we can write explicitly the binary system energy of circular
orbits
\be\label{eq:Ecirclog}
E_{circ}=-\frac 12\eta Mx^2\pag{1+\ldots+x^4\eta\log x
\paq{\frac{448}{15}+x\pa{-\frac{4988}{35}-\frac{656}5\eta}}}\,,
\ee
where we have omitted non-logarithmic terms, completely known up to 4PN order.
The 4PN logarithmic correction was first computed in
\cite{Blanchet:2010zd} and later confirmed with EFT methods in
\cite{Foffa:2011np,Goldberger:2012kf,Galley:2015kus},
and the 5PN computed here agrees with the one found in \cite{LeTiec:2011ab},
later confirmed in \cite{Bini:2013rfa},
by comparison with extreme mass ratio results, i.e. with a non PN computation.

\subsubsection{Angular momentum ``failed'' tail}
\allowdisplaybreaks
The results of the two bottom amplitudes of fig.~\ref{fig:hereditary} are
grouped in literature under the label {\it memory} terms, but, as explained in
sec.~\ref{sec:intro}, we find that both of them actually give
finite, local-in-time contributions to the conservative dynamics and no
contribution to the dissipative one.

In particular the diagram involving the conserved total angular momentum
$\vec{L}$ can be dubbed as a ``failed" tail because, despite having an identical diagrammatic representation to the
energy tail, after replacing $E\to \vec{L}$, and
also being characterized by $q_0=0$, it gives just an instantaneous
contribution to the conservative dynamics.

Indeed, for this diagram at least one of the
graviton polarizations must be an $\vec{A}$ since it is the only polarization
directly coupling to the angular momentum of the system, and it presents
a gradient coupling proportional to momentum $\Q$ that kills any divergence
of the amplitude, which in the case $q_0=0$ occurs for $\Q\to 0$.
We can thus infer that all diagrams involving the conserved quantity $\vec{L}$ and any higher multipole are qualitatively different from the ones involving $E$ in that the former are real, finite and local.

Broken in terms of the polarization, the amplitude for the bottom right
diagram of fig.~\ref{fig:hereditary} is
\be
\label{eq:failTailL}
\ba{rll}
\ds iS_{eff5PN}^{LQ^2}&\ds =-i 64 \pi^2 G^2_d L_p\epsilon_{pmn}
\int\frac{{\rm d}k_0}{2\pi}Q_{ij}(k_0)\dot{Q}_{kl}(-k_0)
\intkq \frac 1{\K^2-k_0^2}\frac 1{(\K+\Q)^2-k_0^2}&\!\!\ds\frac{q^n}{\Q^2}\times\\
&\ds\left\{\frac12 k_0^4 \delta^{kj}\left(\delta^{im}k^l-\delta^{lm}k^i\right)
\right.&\!\!\pag{A\sigma^2}\\
&\ds-\frac12k_0^2 \delta_{lm} q^k k^i k^j&\!\!\{A\sigma\phi\}\\
&\ds+\frac1{2 c_d}(k+q)^i (k+q)^j k^k k^l k^m&\!\!\{A\phi^2\}\\
&\ds-\frac{1}2 k^i k^j \left({\bf q}\cdot({\bf k+q})\delta_{ml}+q^mk^l-q^lk^m\right)(k+q)^k&\!\!\{A^2\phi\}\\
&\ds+\frac12 k_0^2 k^i \left({\bf k}\cdot{\bf q}\delta_{mk}\delta_{jl}+q^k k^l\delta_{mj}+\delta^{mk}(k^j q^l-k^l q^j)+\delta^{jl}(q^m k^k-k^m q^k)\right)&
\!\!\{A^2\sigma\}\\
&\left.\ds-\frac12k_0^2 k^k (k+q)^i\pa{\delta_{lj}k^m-\delta_{mj}k^l+\delta_{lm}k^j}\right\}&\!\!\pag{A^3}\,,
\ea
\ee
which summed up and written in time domain is
\be
\label{eq:failTailL_tot}
S_{eff5PN}^{LQ^2}=\ds \frac{8}{15} G_N^2
\int {\rm d}t\ddddot{Q}_{il}\dddot {Q}_{jl}\epsilon_{ijk}L_k\,.
\ee

\subsubsection{GW self interaction}
\allowdisplaybreaks
Finally we consider the last diagram, which is qualitatively different from the 
others studied so far because it involves three mass quadrupoles, which are
not conserved quantities, and a triple GW vertex.
This is usually considered as a memory term for its effect on the gravitational
waveform, see e.g.~\cite{Blanchet:1997ji}, but it appears as a local-in-time
contribution to the energy.
The detail of the amplitude is
\be
\ba{rll}
iS^{Q^3}_{eff\,5PN}&\ds =\ds -i128\pi^2G^2_d
\int\frac{{\rm d}q_0}{2\pi}\frac{{\rm d}k_0}{2\pi}
Q_{ij}(k_0)Q_{mn}(q_0)Q_{kl}(p_0)\times &\\
&\ds\intkq \frac 1{\K^2-k_0^2}\frac 1{{\bf p}^2-p_0^2}\frac 1{\Q^2-q_0^2}
\times&\\
&\quad\quad\quad\ds\frac13\left[V^{ijmnkl}\left(k,q,p\right)+V^{ijmnkl}\left(k,p,q\right)+V^{ijmnkl}\left(p,q,k\right)\right]&\,,
\ea
\ee
with $p^{\mu}\equiv-k^{\mu}-q^{\mu}$, retarded (advanced) boundary conditions are
used for $k,q$ ($p$),  and
\be\label{eq:QQQret}
\ba{rcll}
V^{ijmnkl}(k,q,p)&=&\ds\left\{
-\frac{p_0^2 k_0^2 q_0^2}8\delta_{im}\delta_{jk}\delta_{ln}\left(-k_0 q_0+{\bf k}\cdot{\bf q}\right)\right.&\{ \sigma^3\}\\
&&\ds-\frac{p_0^3q_0k_0^2}4
q_n\delta_{jl}\pa{k_k\delta_{im}-k_i\delta_{km}}&
\{\sigma^2 A\}\\
&&\ds+0&\{\sigma^2 \phi\}\\
&&\ds-\frac14\delta^{jl}p_0^2 k_0 k_i q_m q_n\left(q_0 k_k-k_0 q_k\right)&\{\sigma A\phi\}\\
&&\ds+\frac18 p_0^2 k_0 q_0 k_i q_m\left[(\K\cdot\Q -k_0 q_0)\delta_{nk}\delta_{jl}+k_l q_k\delta_{nj}+\right.&\\
  &&\ds\left.\delta_{nl}\pa{k_j q_k-k_k q_j}+\delta_{jk}\pa{k_l q_n-k_n q_l}\right]&\{\sigma A^2\}\\
&&\ds-\frac1{8 c_d} p_0^2 k_i k_j k_k q_l q_m q_n&\{\sigma\phi^2\}\\
&&\ds-\frac14 p_0^2 k_0 q_0 k_i p_k q_m \left(q_j\delta_{nl}-q_l\delta_{nj}+q_n\delta_{lj}\right)&\{A^3\}\\
&&\ds\,+\frac18 k_0 p_0k_ip_k q_m q_n \left({\bf k}\cdot{\bf p}\ \delta_{jl}+k_jp_l-k_lp_j\right)&\{A^2\phi\}\\
&&\ds\,+\frac1{4 c_d}k_0 p_0k_i k_j p_k q_l q_m q_n&\{A\phi^2\}\\
&&\left.\ds\,-\frac1{8c_d}k_0 p_0 k_i k_j p_k p_l q_m q_n\right\}&\{\phi^3\}\,.
\ea
\ee

Some of the polarizations above give a divergent result due to the presence of
the master integral $I_m$, but poles cancel in the sum, which can be compactly
written in time domain as in the previous case:
\be
\label{eq:mem}
S^{Q^3}_{eff\,5PN}&=&\ds -\frac{G_N^2}{15}
\int {\rm d}t \left[\ddddot{Q}_{il}\ddddot {Q}_{jl}{Q}_{ij}+\frac4{7}\dddot{Q}_{il}\dddot {Q}_{jl}\ddot{Q}_{ij}\right]\,.
\ee

This concludes our derivation of the hereditary terms at next-to-leading
order for a gravitational source described by the multipolar expansion.
With the exception of eq.~(\ref{eq:Ecirclog}), the
validity of all the results of this subsection is not restricted to the compact
binary case, but it rather holds for any source which allows a multipole
decomposition.

\section{Conclusions}
\label{sec:discussion}
Given the advent of Gravitational Astronomy, and the planning of 
new gravitational wave detectors, like third generation
ground based \cite{Punturo:2010zz} and space detectors \cite{Audley:2017drz},
able to reach higher signal-to-noise ratios than presently operating LIGO 
\cite{TheLIGOScientific:2014jea} and Virgo \cite{TheVirgo:2014hva},
the study of high precision gravity is becoming an urgent program.

Within the post-Newtonian approximation to General Relativity, which is the
main framework for modeling the signals from coalescing binaries detected so far, it is then of outmost interest to increase our perturbative knowledge of binary
dynamics, which at the moment lies at fourth pertubative post-Newtonian order in the conservative sector (see 
\cite{Foffa:2012rn,Foffa:2019rdf,Foffa:2016rgu,Levi:2011eq,Levi:2015uxa,Levi:2015ixa,Levi:2014gsa,Levi:2016ofk,Porto:2010tr,Porto:2008tb,Porto:2008jj}
for a complete determination of the 4PN near zone dynamics purely within the
effective field theory methods),
as well as to gain insight on generic properties of the PN series.

The effective field theory of gravity program, initiated in
\cite{Goldberger:2004jt}, has been proved very powerful in addressing this problem
and within its framework we have derived in the present
paper additional bricks concurring to the edification of the complete fifth
post-Newtonian order binary dynamics.

In particular at 5PN, like at 4PN, there are contributions from the far, or
radiation zone, where the degrees of freedom of gravity couple to
source multipoles, to the near zone dynamics, i.e. the region around the
source whose size is smaller than the wavelength of gravitational waves.
The division in zones leads to several operational simplifications within the
post-Newtonian approximation but also introduces spurious divergences in both
zones starting from 4PN order, which
recompose to a finite physical result once the two computations are consistently
combined, as explained in detail in \cite{Foffa:2019yfl}.

However computable, finite local-in-time and unambiguous terms remain after near and far zone
results are combined, and  
we have originally derived in the present paper all yet unknown contributions
from the far zone to the near zone conservative dynamics at 5PN order,
which we report here for convenience of the reader

\be\label{eq:magicnumbers}
S^{(far)}_{eff\,5PN}=G_N^2 \int {\rm d} t&&
\left[M\pa{\frac{82}{6615}\ddddot{O}_{ijk}^2
+\frac{508}{675}\dddot{J}_{ij}^2}\right.\nonumber\\
&&\left.+\frac8{15}\ddddot{Q}_{il}\dddot {Q}_{jl}\epsilon_{ijk}L_k-\frac1{15}\ddddot{Q}_{il}\ddddot {Q}_{jl}Q_{ij}-\frac4{105}\dddot{Q}_{il}\dddot {Q}_{jl}\ddot{Q}_{ij}\right]\,.
\ee
In particular, the terms in the second line come from finite, local amplitudes
in the far zone, so there are no associated IR-divergent terms in the near zone
that could possibly signal the presence of such finite contributions.

Along with the finite local-in-time terms, there comes logarithmic hereditary terms, whose values we
obtained in agreement with known previous results \cite{LeTiec:2011ab,Bini:2013rfa}, which have been extended in \cite{Blanchet:2019rjs}.

Note that the 5th PN order is qualitative different from previous ones, since
it is the lowest one at which finite size effects, for spin-less black holes,
are not forbidden by the effacement principle \cite{Damour:1982wm},
even though they are expected to appear only at higher order because of the
vanishing of black hole static Love number \cite{Binnington:2009bb,Damour:2009vw,Kol:2011vg,Gurlebeck:2015xpa,Pani:2015hfa,Pani:2015nua,Landry:2015zfa}.

To determine the remaining missing terms ruling the 5PN dynamics it is necessary
to complete the near zone computations, which has already been solved in the
static sector, i.e. at $O(G_N^6)$, in \cite{Foffa:2019hrb} (independently confirmed in
\cite{Blumlein:2019zku}), while it is in principle possible to extract information about the 5PN order
at lower power of $G_N$ from the post-Minkowskian results at $O(G_N)$
\cite{Westpfahl:1979gu,Ledvinka:2008tk,Foffa:2013gja,Damour:2016gwp,Blanchet:2018yvb}, $O(G_N^2)$ \cite{Damour:2017zjx,Cheung:2018wkq}, $O(G_N^3)$ \cite{Bern:2019nnu}.
Among all the terms needed at 5PN, the ones determined in this paper stand out as the only ones that require knowledge of the far zone dynamics.

As a byproduct of our computation we have also re-derived the universality relations, already observed in \cite{Blanchet:1995fr},
between the power flux emitted by any multipole moment via the tail process
and the leading order.

Finally the last original finding of the present work has been to relate the
flux formula at generic $n$-PN order to the logarithm of tail terms
affecting the real part of the action at $(n+4)$-PN order, as summarized by
eq.~(\ref{eq:tail_univ}).
These logarithmic terms embody non-local-in-time (but causal)
interactions depending on the past history of the source, which become
instantaneous (hence local) for multipoles describing binary systems in circular
orbit.

The logarithmic terms come with (unphysical) poles, which then are also
constrained by the flux emission formula. Since the far zone poles has to cancel
with equally unphysical poles in the near zone, this provides another
non-trivial constraints on the results of
near zone dynamics that are needed to complete the 5PN order dynamics.

\section*{Acknowledgments}
R.~S. is partially supported by CNPq.
S.~F. is supported by the Fonds National Suisse and by the SwissMAP NCCR National Center of Competence in Research.
The authors wish to thank Gabriel Luz Almeida for checking the
computations of the manuscript and correcting a typo in one of the equations.

\appendix

\section{Amplitude construction}
\label{app:sample_ampl}

To construct amplitudes the basic building blocks are the multipolar action
(\ref{eq:mult}) and the gauge-fixed and bulk gravity action (\ref{eq:sEH_KK}).
E.g.~to derive (\ref{eq:PQQ}) one has to consider the quadrupole-gravity linear
coupling, from the explicit expression eq.~(\ref{eq:riem_el}), which in Fourier
domain is written as
\be
\label{eq:QRfourier}
\int \frac{{\rm d}^{d+1}k}{(2\pi)^{d+1}}\frac{Q_{ij}(k_0)}{2\Lambda}
\paq{-\frac 12k_0^2 \sigma_{ij}-\frac 12k_0k_iA_j-\frac 12k_0k_jA_i+
  \pa{k_ik_j+\frac{k_0^2}{d-2}\delta_{ij}}\phi}\,,
\ee
and the gravity propagators
\be
\label{eq:props}
\left.\ba{lcl}
\ds P[\sigma_{ij}(k)\sigma_{kl}(q)]&=&\ds -\pa{\delta_{ik}\delta_{jl}+\delta_{il}\delta_{jk}-\frac 2{d-2}\delta_{ij}\delta_{kl}}\\
\ds P[A_i(k)A_j(q)]&=&\delta_{ij}\\
\ds P[\phi(k)\phi(q)]&=&\ds -\frac{d-2}{2(d-1)}
\ea\right\}\times i\frac{(2\pi)^{d+1}\delta^{(d+1)}(k+q)}{2(\K^2-k_0^2)}\,,
\ee
with vanishing propagators between not-alike fields (i.e. $P[\sigma_{ij}A_k]=P[\sigma_{ij}\phi]=P[A_i\phi]=0$).

Gluing together two terms of the type of (\ref{eq:QRfourier}), using the
above propagators (\ref{eq:props}) and integrating over all possible momenta
one finally obtains eq.~(\ref{eq:PQQ}).

To build tail amplitudes one has to consider a tri-linear bulk coupling, which
can be read from (\ref{eq:sEH_KK}), and pair each of its three fields with a
source multipole term.
E.g.~to build the amplitude corresponding to the first line in the curly bracket of eq.~(\ref{eq:tailQQ}) one has to consider 
bulk vertex
\be
\frac{c_dk_0k'_0}{4\Lambda}\pa{\sigma_{ij}(k)\sigma_{kl}(k')\phi(q)}
\pa{\delta_{ik}\delta_{jl}+\delta_{il}\delta_{jk}-\delta_{ij}\delta_{kl}}\,,
\ee
then Wick-contract each $\phi$ and $\sigma$ field with the the appropriate field in
\bdm
\paq{\int \frac{d^{d+1}k}{(2\pi)^{d+1}}\frac{(-k_0^2)}4Q_{ij}(k_0)\sigma_{ij}(-k)}^2
\pa{\int \frac{d^{d+1}q}{(2\pi)^{d+1}}\frac{(-E)}\Lambda\phi(q)}\,,
\edm
where in the propagator connected with the source term involving the conserved energy $E$ the denominator has to be expanded for small $q_0$, i.e.
\be
\label{eq:region}
\frac 1{\Q^2-q_0^2}\simeq \frac 1{\Q^2}\pa{1+\frac{q_0^2}{\Q^2}+\ldots}\,,
\ee
since the momentum of the longitudinal gravitational mode $q=(q_0,\Q)$ has
$q_0\ll |\Q|$, according to the integration prescription known as region of
momentum (see \cite{Jantzen:2011nz} for a rigorous demonstration), and at
leading order all terms involving $q_0$ in eq.~(\ref{eq:region}) can be
neglected.

\section{Relevant integrals}
\label{app:integralacci}
In the present work we had to integrate amplitudes like eq.(\ref{eq:her_gen}), involving numerators with up to six free indices;
exploiting spatial rotation invariance and the possibility of relating integrals by means of index contractions, everything can be reduced to the
integration of the following scalar factor
\be
\frac1{{\cal D}_{\alpha,\beta,\gamma}}\equiv\paq{\pa{\Q^2-q_0^2}^\alpha\pa{(\K+\Q)^2-\pa{k_0+q_0}^2}^\beta\pa{\K^2-k_0^2}^\gamma}^{-1}\,,
\ee
with $(\alpha, \beta,\gamma)$ integers equal to $-2$, $-1$, $0$ or $1$ and the
retarded prescription for propagator poles is understood.
The case $\alpha=\beta=\gamma=1$ corresponds to the master integral $I_m$, while
the other relevant cases, using
\be
\label{eq:int0}
\int_\K\frac 1{\K^2-k_0^2}=\frac{\Gamma(1-d/2)}{(4\pi)^{d/2}}(-k_0^2)^{d/2-1}\,,
\ee
give
\be
\ba{rcl}
\ds\intkq \frac1{{\cal D}_{1,0,1}}&=&\ds\Theta (-k_0^2)^{d/2-1}(-q_0^2)^{d/2-1}\,,
\quad\quad\Theta\equiv\frac{\Gamma^2\left(1-\frac d2\right)}{(4 \pi)^d}\,,\\
\ds\intkq \frac1{{\cal D}_{-n,0,\alpha}}&=&\ds\intkq \frac1{{\cal D}_{\alpha,0,-n}}=0 \quad {\rm for}\quad n\geq 0\,,\\
\ds\intkq \frac1{{\cal D}_{1,-1,1}}&=&\ds\paq{q_0^2+k_0^2-\pa{k_0+q_0}^2}\intkq \frac1{{\cal D}_{1,0,1}}\,,\\
\ds\intkq \frac1{{\cal D}_{1,-2,1}}&=&\ds\left[(q_0^2+k_0^2)^2-\pa{k_0+q_0}^4+\frac4d k_0^2 q_0^2)\right]\intkq \frac1{{\cal D}_{1,0,1}}-2 \pa{k_0+q_0}^2\intkq \frac1{{\cal D}_{1,-1,1}}\,.
\ea
\ee
The results for the integrals with two free indices can be written, up to terms ${\cal O}(d-3)$, as
\be
\ba{rcl}
\ds\int_\K\frac{k^ik^j}{\K^2-k_0^2}&=&\ds\frac{\delta^{ij}k_0^2}d
\frac{\Gamma(1-d/2)}{(4\pi)^{d/2}}\pa{-k_0^2}^{d/2-1}\,,\\
\ds\int_{\K,\Q}k^iq^j\frac1{\cal D}&\simeq&\ds\frac{\delta^{ij}}d\paq{ I_mq_0k_0+\frac{\Theta}2\pa{k_0^2+q_0^2+k_0q_0}}\,,\\
\ds\int_{\K,\Q}k^ik^j\frac1{\cal D}&\simeq&\ds\frac{\delta^{ij}}d\paq{ I_m k_0^2-\Theta q_0\pa{k_0+q_0}}\,,\\
\ds\int_{\K,\Q}q^iq^j\frac1{\cal D}&\simeq&\ds\frac{\delta^{ij}}d\paq{ I_m q_0^2-\Theta k_0\pa{k_0+q_0}}\,.
\ea
\ee
For the four-indices case, in terms of the following parametrization
\be
\ba{cl}
\ds\int k^i k^j k^l k^m \frac1{\cal D}=C^{kkkk}\delta_{ijlm}\,,&
\ds\quad\int k^i k^j k^l q^m \frac1{\cal D}=C^{kkkq}\delta_{ijlm}\,,
\quad\int k^i q^j q^l q^m \frac1{\cal D}=C^{kqqq}\delta_{ijlm}\,,\\
\ds\int q^i q^j q^l q^m \frac1{\cal D}=C^{qqqq}\delta_{ijlm}\,,&\ds
\quad\int k^i k^j q^l q^m \frac1{\cal D}=C^{kkqq}\delta_{ijlm}+\bar{C}^{kkqq}\delta_{ij}\delta_{lm}\,,
\ea
\ee
where $\delta_{ijlm}\equiv\delta_{il}\delta_{im}+\delta_{im}\delta_{jn}+\delta_{ij}\delta_{lm}$ is the completely symmetrized combination of two $\delta$s,
one obtains
\be
\ba{rcl}
\ds C^{kkkk}&\simeq&\ds\frac1{d(d+2)}\left[k_0^4 I_m-2\Theta q_0 \left(k_0^3+ 2 k_0^2 q_0+2 k_0 q_0^2+q_0^3\right)\right]\,,\\
\ds C^{kkkq}&\simeq&\ds\frac1{d(d+2)}\left[k_0^3 q_0I_m+\frac{\Theta}2 \left(k_0^4+ k_0^3 q_0+k_0^2 q_0^2+2 k_0 q_0^3+2q_0^4\right)\right]\,,\\
\ds C^{kkqq}&\simeq&\ds\frac1{d(d+2)}\left\{k_0^2 q_0^2 I_m-\frac1{2(d-1)}\Theta \left[(d-2)k_0^4-2k_0^3 q_0-dk_0^2 q_0^2-2k_0 q_0^3+(d-2)q_0^4\right]\right\}\,,\\
\ds \bar{C}^{kkqq}&\simeq&\ds-\frac{\Theta}{2d(d-1)} \left[k_0^4+ 2k_0^3 q_0+k_0^2 q_0^2+ 2k_0 q_0^3+q_0^4\right]\,.
\ea
\ee
and the results for $C^{qqqq}$ and  $C^{kqqq}$ can obviously be obtained from $C^{kkkk}$ and  $C^{kkkq}$ by means of $k_0\leftrightarrow q_0$.

Finally, by using an analogous parametrization for the six-indices integrals
\be
&&\int k^i k^j k^l k^m k^r k^s \frac1{\cal D}=C^{kkkkkk}\delta_{ijlmrs}\,,\quad\int k^i k^j k^l k^m k^r q^s \frac1{\cal D}=C^{kkkkkq}\delta_{ijlmrs}\,,\nonumber\\
&&\int k^i q^j q^l q^m q^r q^s \frac1{\cal D}=C^{kqqqqq}\delta_{ijlmrs}\,,\quad\int q^i q^j q^l q^m q^r q^s \frac1{\cal D}=C^{qqqqqq}\delta_{ijlmrs}\,,\nonumber\\
&&\int k^i k^j k^l k^m q^r q^s \frac1{\cal D}=C^{kkkkqq}\delta_{ijlmrs}+\bar{C}^{kkkkqq}\delta_{ijlm}\delta_{rs}\,,\nonumber\\
&&\int k^i k^j q^l q^m q^r q^s \frac1{\cal D}=C^{kkqqqq}\delta_{ijlmrs}+\bar{C}^{kkqqqq}\delta_{lmrs}\delta_{ij}\,,\nonumber\\
&&\int k^i k^j k^l q^m q^r q^s \frac1{\cal D}=C^{kkkqqq}\delta_{ijlmrs}\nonumber\\
&&\quad\quad\quad\quad\quad\quad\quad\quad+\bar{C}^{kkkqqq}\left[\delta_{im}\left(\delta_{jr}\delta_{ls}+\delta_{js}\delta_{lr}\right)+\delta_{ir}\left(\delta_{jm}\delta_{ls}+\delta_{js}\delta_{lm}\right)+\delta_{is}\left(\delta_{jm}\delta_{lr}+\delta_{jr}\delta_{lm}\right)\right]\,,\nonumber
\ee
where $\delta_{ijlmrs}$ is the completely symmetrized combination of three $\delta$'s, one gets
\be
C^{kkkkkk}&\simeq&  \frac{k_0^6 I_m-\frac{\Theta}d q_0 \left[3 d k_0^5+ 9 d k_0^4 q_0+4(4d+1)k_0^3 q_0^2 +6(3d +2)k_0^2 q_0^3+12(d+1) k_0 q_0^4+4(d+1)q_0^5\right]}{d(d+2)(d+4)}\nonumber\\
C^{kkkkkq}&\simeq&  \frac{k_0^5 q_0 I_m+\frac{\Theta}{2d}  \left[d k_0^6 +d k_0^5 q_0+d k_0^4 q_0^2 +4(d+1)k_0^3 q_0^3 +4(2d +3)k_0^2 q_0^4+4(2d+3) k_0 q_0^5+4(d+1)q_0^6\right]}{d(d+2)(d+4)}\nonumber\\
C^{kkkkqq}&\simeq&  \frac{k_0^4 q_0^2I_m-\frac{\Theta}{2(d-1)}\left[(d-2)k_0^6 -2k_0^5 q_0-d k_0^4 q_0^2 -2k_0^3 q_0^3+(d-2)k_0^2 q_0^4+2(d-1) k_0 q_0^5+2(d-1)q_0^6\right]}{d(d+2)(d+4)}\nonumber\\
\bar{C}^{kkkkqq}&\simeq&-\Theta\frac{d k_0^6 +2 d k_0^5 q_0+d k_0^4 q_0^2+2(2d-1)k_0^3 q_0^3+(7d-6)k_0^2 q_0^4+6(d-1) k_0 q_0^5+2(d-1)q_0^6}{2d^2(d-1)(d+2)}\nonumber\\
C^{kkkqqq}&\simeq&  \frac{k_0^3q_0^3}{d(d+2)(d+4)} I_m+\frac{\Theta}{2d^2(d-1)(d+2)(d+4)}\left[2(d^2-1)k_0^6 +2(d^2+d-3)k_0^5 q_0 \right.\nonumber\\
&&\left.+(d+3)(d-2)k_0^4 q_0^2+(d-2)(d+1)k_0^3 q_0^3+(d+3)(d-2)k_0^2 q_0^4+2(d^2+d-\!3) k_0 q_0^5+2(d^2\!-\!1)q_0^6\!\right]\nonumber\\
\bar{C}^{kkkqqq}&\simeq&  -\Theta\frac{(d-1)k_0^6 +(2d-3)k_0^5 q_0+(d-3)k_0^4 q_0^2 -k_0^3 q_0^3+(d-3)k_0^2 q_0^4+(2d-3) k_0 q_0^5+(d-1)q_0^6}{2d^2(d-1)(d+2)}\,,\nonumber\\
\ee
and the results for $C^{qqqqqq}$, $C^{kqqqqq}$, $C^{kkqqqq}$ and $\bar{C}^{kkqqqq}$ can be obtained as above by means of $k_0\leftrightarrow q_0$.

As in most cases one can set $q_0=0$ because a conserved quantity is involved in the amplitude, one can exploit the following closed formula
\be
&&\intkq \left.\frac1{{\cal D}_{\alpha,\beta,\gamma}}\right|_{q_0=0}=
\ds
\frac{\Gamma(\alpha+\beta+\gamma-d)\Gamma(\alpha+\gamma-\frac d2)\Gamma(\alpha+\beta-\frac d2)\Gamma(\frac d2-\alpha)}{(4\pi)^d\Gamma(\beta)\Gamma(\gamma)\Gamma(2\alpha+\beta+\gamma-d)\Gamma(\frac d2)}
\pa{-k_0^2}^{d-\alpha-\beta-\gamma}\,.
\nonumber
\ee

The dimension-less function $f$ appearing in eq.~(\ref{eq:tailf}) is
  explicitly given by $k_0^2(-k_0^2)^{1-d/2}$ times
\be
\ba{rcl}
\ds\int_\Q\frac 1{\Q^2[(\K+\Q)^2-k_0^2]}&=&\ds\frac{\Gamma(d/2-1)}{(4\pi)^{d/2}}
\int_0^1 dx \pa{x(1-x)\K^2-xk_0^2}^{d/2-2}\\
&=&\ds -\frac{2\Gamma(d/2-1)}{(d-2)(4\pi)^{d/2}}
\frac{(-k_0^2)^{d/2-1}}{\K^2-k_0^2}
{}_2F_1\pa{1,d-2;\frac d2;\frac{\K^2}{\K^2-k_0^2}}\,,
\ea
\ee
where ${}_2F_1(a,b,c;z)$ is the Hypergeometric function which is divergent for
$|z|\to 0$.


\begin{thebibliography}{89}%
\makeatletter
\providecommand \@ifxundefined [1]{%
 \@ifx{#1\undefined}
}%
\providecommand \@ifnum [1]{%
 \ifnum #1\expandafter \@firstoftwo
 \else \expandafter \@secondoftwo
 \fi
}%
\providecommand \@ifx [1]{%
 \ifx #1\expandafter \@firstoftwo
 \else \expandafter \@secondoftwo
 \fi
}%
\providecommand \natexlab [1]{#1}%
\providecommand \enquote  [1]{``#1''}%
\providecommand \bibnamefont  [1]{#1}%
\providecommand \bibfnamefont [1]{#1}%
\providecommand \citenamefont [1]{#1}%
\providecommand \href@noop [0]{\@secondoftwo}%
\providecommand \href [0]{\begingroup \@sanitize@url \@href}%
\providecommand \@href[1]{\@@startlink{#1}\@@href}%
\providecommand \@@href[1]{\endgroup#1\@@endlink}%
\providecommand \@sanitize@url [0]{\catcode `\\12\catcode `\$12\catcode
  `\&12\catcode `\#12\catcode `\^12\catcode `\_12\catcode `\%12\relax}%
\providecommand \@@startlink[1]{}%
\providecommand \@@endlink[0]{}%
\providecommand \url  [0]{\begingroup\@sanitize@url \@url }%
\providecommand \@url [1]{\endgroup\@href {#1}{\urlprefix }}%
\providecommand \urlprefix  [0]{URL }%
\providecommand \Eprint [0]{\href }%
\providecommand \doibase [0]{http://dx.doi.org/}%
\providecommand \selectlanguage [0]{\@gobble}%
\providecommand \bibinfo  [0]{\@secondoftwo}%
\providecommand \bibfield  [0]{\@secondoftwo}%
\providecommand \translation [1]{[#1]}%
\providecommand \BibitemOpen [0]{}%
\providecommand \bibitemStop [0]{}%
\providecommand \bibitemNoStop [0]{.\EOS\space}%
\providecommand \EOS [0]{\spacefactor3000\relax}%
\providecommand \BibitemShut  [1]{\csname bibitem#1\endcsname}%
\let\auto@bib@innerbib\@empty
\bibitem [{\citenamefont {Abbott}\ \emph
  {et~al.}(2019{\natexlab{a}})\citenamefont {Abbott} \emph
  {et~al.}}]{LIGOScientific:2018mvr}%
  \BibitemOpen
  \bibfield  {author} {\bibinfo {author} {\bibfnamefont {B.~P.}\ \bibnamefont
  {Abbott}} \emph {et~al.} (\bibinfo {collaboration} {LIGO Scientific,
  Virgo}),\ }\href {\doibase 10.1103/PhysRevX.9.031040} {\bibfield  {journal}
  {\bibinfo  {journal} {Phys. Rev.}\ }\textbf {\bibinfo {volume} {X9}},\
  \bibinfo {pages} {031040} (\bibinfo {year} {2019}{\natexlab{a}})},\ \Eprint
  {http://arxiv.org/abs/1811.12907} {arXiv:1811.12907 [astro-ph.HE]}
  \BibitemShut {NoStop}%
\bibitem [{\citenamefont {Aasi}\ \emph {et~al.}(2015)\citenamefont {Aasi} \emph
  {et~al.}}]{TheLIGOScientific:2014jea}%
  \BibitemOpen
  \bibfield  {author} {\bibinfo {author} {\bibfnamefont {J.}~\bibnamefont
  {Aasi}} \emph {et~al.} (\bibinfo {collaboration} {LIGO Scientific}),\ }\href
  {\doibase 10.1088/0264-9381/32/7/074001} {\bibfield  {journal} {\bibinfo
  {journal} {Class. Quant. Grav.}\ }\textbf {\bibinfo {volume} {32}},\ \bibinfo
  {pages} {074001} (\bibinfo {year} {2015})},\ \Eprint
  {http://arxiv.org/abs/1411.4547} {arXiv:1411.4547 [gr-qc]} \BibitemShut
  {NoStop}%
\bibitem [{\citenamefont {Acernese}\ \emph {et~al.}(2015)\citenamefont
  {Acernese} \emph {et~al.}}]{TheVirgo:2014hva}%
  \BibitemOpen
  \bibfield  {author} {\bibinfo {author} {\bibfnamefont {F.}~\bibnamefont
  {Acernese}} \emph {et~al.} (\bibinfo {collaboration} {VIRGO}),\ }\href
  {\doibase 10.1088/0264-9381/32/2/024001} {\bibfield  {journal} {\bibinfo
  {journal} {Class. Quant. Grav.}\ }\textbf {\bibinfo {volume} {32}},\ \bibinfo
  {pages} {024001} (\bibinfo {year} {2015})},\ \Eprint
  {http://arxiv.org/abs/1408.3978} {arXiv:1408.3978 [gr-qc]} \BibitemShut
  {NoStop}%
\bibitem [{\citenamefont {Allen}\ \emph {et~al.}(2012)\citenamefont {Allen},
  \citenamefont {Anderson}, \citenamefont {Brady}, \citenamefont {Brown},\ and\
  \citenamefont {Creighton}}]{Allen:2005fk}%
  \BibitemOpen
  \bibfield  {author} {\bibinfo {author} {\bibfnamefont {B.}~\bibnamefont
  {Allen}}, \bibinfo {author} {\bibfnamefont {W.~G.}\ \bibnamefont {Anderson}},
  \bibinfo {author} {\bibfnamefont {P.~R.}\ \bibnamefont {Brady}}, \bibinfo
  {author} {\bibfnamefont {D.~A.}\ \bibnamefont {Brown}}, \ and\ \bibinfo
  {author} {\bibfnamefont {J.~D.~E.}\ \bibnamefont {Creighton}},\ }\href
  {\doibase 10.1103/PhysRevD.85.122006} {\bibfield  {journal} {\bibinfo
  {journal} {Phys. Rev.}\ }\textbf {\bibinfo {volume} {D85}},\ \bibinfo {pages}
  {122006} (\bibinfo {year} {2012})},\ \Eprint
  {http://arxiv.org/abs/gr-qc/0509116} {arXiv:gr-qc/0509116 [gr-qc]}
  \BibitemShut {NoStop}%
\bibitem [{\citenamefont {Blanchet}(2014)}]{Blanchet:2013haa}%
  \BibitemOpen
  \bibfield  {author} {\bibinfo {author} {\bibfnamefont {L.}~\bibnamefont
  {Blanchet}},\ }\href {\doibase 10.12942/lrr-2014-2} {\bibfield  {journal}
  {\bibinfo  {journal} {Living Rev. Rel.}\ }\textbf {\bibinfo {volume} {17}},\
  \bibinfo {pages} {2} (\bibinfo {year} {2014})},\ \Eprint
  {http://arxiv.org/abs/1310.1528} {arXiv:1310.1528 [gr-qc]} \BibitemShut
  {NoStop}%
\bibitem [{\citenamefont {Mrou\'e}\ \emph {et~al.}(2013)\citenamefont {Mrou\'e}
  \emph {et~al.}}]{Mroue:2013xna}%
  \BibitemOpen
  \bibfield  {author} {\bibinfo {author} {\bibfnamefont {A.~H.}\ \bibnamefont
  {Mrou\'e}} \emph {et~al.},\ }\href {\doibase 10.1103/PhysRevLett.111.241104}
  {\bibfield  {journal} {\bibinfo  {journal} {Phys. Rev. Lett.}\ }\textbf
  {\bibinfo {volume} {111}},\ \bibinfo {pages} {241104} (\bibinfo {year}
  {2013})},\ \Eprint {http://arxiv.org/abs/1304.6077} {arXiv:1304.6077 [gr-qc]}
  \BibitemShut {NoStop}%
\bibitem [{\citenamefont {Taracchini}\ \emph {et~al.}(2014)\citenamefont
  {Taracchini} \emph {et~al.}}]{Taracchini:2013rva}%
  \BibitemOpen
  \bibfield  {author} {\bibinfo {author} {\bibfnamefont {A.}~\bibnamefont
  {Taracchini}} \emph {et~al.},\ }\href {\doibase 10.1103/PhysRevD.89.061502}
  {\bibfield  {journal} {\bibinfo  {journal} {Phys. Rev.}\ }\textbf {\bibinfo
  {volume} {D89}},\ \bibinfo {pages} {061502} (\bibinfo {year} {2014})},\
  \Eprint {http://arxiv.org/abs/1311.2544} {arXiv:1311.2544 [gr-qc]}
  \BibitemShut {NoStop}%
\bibitem [{\citenamefont {Pan}\ \emph {et~al.}(2014)\citenamefont {Pan},
  \citenamefont {Buonanno}, \citenamefont {Taracchini}, \citenamefont {Kidder},
  \citenamefont {Mrou\'e}, \citenamefont {Pfeiffer}, \citenamefont {Scheel},\
  and\ \citenamefont {Szil\'agyi}}]{Pan:2013rra}%
  \BibitemOpen
  \bibfield  {author} {\bibinfo {author} {\bibfnamefont {Y.}~\bibnamefont
  {Pan}}, \bibinfo {author} {\bibfnamefont {A.}~\bibnamefont {Buonanno}},
  \bibinfo {author} {\bibfnamefont {A.}~\bibnamefont {Taracchini}}, \bibinfo
  {author} {\bibfnamefont {L.~E.}\ \bibnamefont {Kidder}}, \bibinfo {author}
  {\bibfnamefont {A.~H.}\ \bibnamefont {Mrou\'e}}, \bibinfo {author}
  {\bibfnamefont {H.~P.}\ \bibnamefont {Pfeiffer}}, \bibinfo {author}
  {\bibfnamefont {M.~A.}\ \bibnamefont {Scheel}}, \ and\ \bibinfo {author}
  {\bibfnamefont {B.}~\bibnamefont {Szil\'agyi}},\ }\href {\doibase
  10.1103/PhysRevD.89.084006} {\bibfield  {journal} {\bibinfo  {journal} {Phys.
  Rev.}\ }\textbf {\bibinfo {volume} {D89}},\ \bibinfo {pages} {084006}
  (\bibinfo {year} {2014})},\ \Eprint {http://arxiv.org/abs/1307.6232}
  {arXiv:1307.6232 [gr-qc]} \BibitemShut {NoStop}%
\bibitem [{\citenamefont {Hannam}\ \emph {et~al.}(2014)\citenamefont {Hannam},
  \citenamefont {Schmidt}, \citenamefont {Boh\'e}, \citenamefont {Haegel},
  \citenamefont {Husa}, \citenamefont {Ohme}, \citenamefont {Pratten},\ and\
  \citenamefont {P{\"u}rrer}}]{Hannam:2013oca}%
  \BibitemOpen
  \bibfield  {author} {\bibinfo {author} {\bibfnamefont {M.}~\bibnamefont
  {Hannam}}, \bibinfo {author} {\bibfnamefont {P.}~\bibnamefont {Schmidt}},
  \bibinfo {author} {\bibfnamefont {A.}~\bibnamefont {Boh\'e}}, \bibinfo
  {author} {\bibfnamefont {L.}~\bibnamefont {Haegel}}, \bibinfo {author}
  {\bibfnamefont {S.}~\bibnamefont {Husa}}, \bibinfo {author} {\bibfnamefont
  {F.}~\bibnamefont {Ohme}}, \bibinfo {author} {\bibfnamefont {G.}~\bibnamefont
  {Pratten}}, \ and\ \bibinfo {author} {\bibfnamefont {M.}~\bibnamefont
  {P{\"u}rrer}},\ }\href {\doibase 10.1103/PhysRevLett.113.151101} {\bibfield
  {journal} {\bibinfo  {journal} {Phys. Rev. Lett.}\ }\textbf {\bibinfo
  {volume} {113}},\ \bibinfo {pages} {151101} (\bibinfo {year} {2014})},\
  \Eprint {http://arxiv.org/abs/1308.3271} {arXiv:1308.3271 [gr-qc]}
  \BibitemShut {NoStop}%
\bibitem [{\citenamefont {Goldberger}\ and\ \citenamefont
  {Rothstein}(2006)}]{Goldberger:2004jt}%
  \BibitemOpen
  \bibfield  {author} {\bibinfo {author} {\bibfnamefont {W.~D.}\ \bibnamefont
  {Goldberger}}\ and\ \bibinfo {author} {\bibfnamefont {I.~Z.}\ \bibnamefont
  {Rothstein}},\ }\href {\doibase 10.1103/PhysRevD.73.104029} {\bibfield
  {journal} {\bibinfo  {journal} {Phys. Rev.}\ }\textbf {\bibinfo {volume}
  {D73}},\ \bibinfo {pages} {104029} (\bibinfo {year} {2006})},\ \Eprint
  {http://arxiv.org/abs/hep-th/0409156} {arXiv:hep-th/0409156 [hep-th]}
  \BibitemShut {NoStop}%
\bibitem [{\citenamefont {Goldberger}(2007)}]{Goldberger:2007hy}%
  \BibitemOpen
  \bibfield  {author} {\bibinfo {author} {\bibfnamefont {W.~D.}\ \bibnamefont
  {Goldberger}},\ }in\ \href@noop {} {\emph {\bibinfo {booktitle} {{Les Houches
  Summer School - Session 86: Particle Physics and Cosmology: The Fabric of
  Spacetime Les Houches, France, July 31-August 25, 2006}}}}\ (\bibinfo {year}
  {2007})\ \Eprint {http://arxiv.org/abs/hep-ph/0701129} {arXiv:hep-ph/0701129
  [hep-ph]} \BibitemShut {NoStop}%
\bibitem [{\citenamefont {Foffa}\ and\ \citenamefont
  {Sturani}(2014)}]{Foffa:2013qca}%
  \BibitemOpen
  \bibfield  {author} {\bibinfo {author} {\bibfnamefont {S.}~\bibnamefont
  {Foffa}}\ and\ \bibinfo {author} {\bibfnamefont {R.}~\bibnamefont
  {Sturani}},\ }\href {\doibase 10.1088/0264-9381/31/4/043001} {\bibfield
  {journal} {\bibinfo  {journal} {Class. Quant. Grav.}\ }\textbf {\bibinfo
  {volume} {31}},\ \bibinfo {pages} {043001} (\bibinfo {year} {2014})},\
  \Eprint {http://arxiv.org/abs/1309.3474} {arXiv:1309.3474 [gr-qc]}
  \BibitemShut {NoStop}%
\bibitem [{\citenamefont {Porto}(2016)}]{Porto:2016pyg}%
  \BibitemOpen
  \bibfield  {author} {\bibinfo {author} {\bibfnamefont {R.~A.}\ \bibnamefont
  {Porto}},\ }\href {\doibase 10.1016/j.physrep.2016.04.003} {\bibfield
  {journal} {\bibinfo  {journal} {Phys. Rept.}\ }\textbf {\bibinfo {volume}
  {633}},\ \bibinfo {pages} {1} (\bibinfo {year} {2016})},\ \Eprint
  {http://arxiv.org/abs/1601.04914} {arXiv:1601.04914 [hep-th]} \BibitemShut
  {NoStop}%
\bibitem [{\citenamefont {Levi}(2020)}]{Levi:2018nxp}%
  \BibitemOpen
  \bibfield  {author} {\bibinfo {author} {\bibfnamefont {M.}~\bibnamefont
  {Levi}},\ }\href {\doibase 10.1088/1361-6633/ab12bc} {\bibfield  {journal}
  {\bibinfo  {journal} {Rept. Prog. Phys.}\ }\textbf {\bibinfo {volume} {83}},\
  \bibinfo {pages} {075901} (\bibinfo {year} {2020})},\ \Eprint
  {http://arxiv.org/abs/1807.01699} {arXiv:1807.01699 [hep-th]} \BibitemShut
  {NoStop}%
\bibitem [{\citenamefont {Damour}\ \emph {et~al.}(2015)\citenamefont {Damour},
  \citenamefont {Jaranowski},\ and\ \citenamefont
  {Sch{\"a}fer}}]{Damour:2015isa}%
  \BibitemOpen
  \bibfield  {author} {\bibinfo {author} {\bibfnamefont {T.}~\bibnamefont
  {Damour}}, \bibinfo {author} {\bibfnamefont {P.}~\bibnamefont {Jaranowski}},
  \ and\ \bibinfo {author} {\bibfnamefont {G.}~\bibnamefont {Sch{\"a}fer}},\
  }\href {\doibase 10.1103/PhysRevD.91.084024} {\bibfield  {journal} {\bibinfo
  {journal} {Phys. Rev.}\ }\textbf {\bibinfo {volume} {D91}},\ \bibinfo {pages}
  {084024} (\bibinfo {year} {2015})},\ \Eprint
  {http://arxiv.org/abs/1502.07245} {arXiv:1502.07245 [gr-qc]} \BibitemShut
  {NoStop}%
\bibitem [{\citenamefont {Damour}\ \emph {et~al.}(2016)\citenamefont {Damour},
  \citenamefont {Jaranowski},\ and\ \citenamefont
  {Sch{\"a}fer}}]{Damour:2016abl}%
  \BibitemOpen
  \bibfield  {author} {\bibinfo {author} {\bibfnamefont {T.}~\bibnamefont
  {Damour}}, \bibinfo {author} {\bibfnamefont {P.}~\bibnamefont {Jaranowski}},
  \ and\ \bibinfo {author} {\bibfnamefont {G.}~\bibnamefont {Sch{\"a}fer}},\
  }\href {\doibase 10.1103/PhysRevD.93.084014} {\bibfield  {journal} {\bibinfo
  {journal} {Phys. Rev.}\ }\textbf {\bibinfo {volume} {D93}},\ \bibinfo {pages}
  {084014} (\bibinfo {year} {2016})},\ \Eprint
  {http://arxiv.org/abs/1601.01283} {arXiv:1601.01283 [gr-qc]} \BibitemShut
  {NoStop}%
\bibitem [{\citenamefont {Bernard}\ \emph {et~al.}(2016)\citenamefont
  {Bernard}, \citenamefont {Blanchet}, \citenamefont {Boh\'e}, \citenamefont
  {Faye},\ and\ \citenamefont {Marsat}}]{Bernard:2015njp}%
  \BibitemOpen
  \bibfield  {author} {\bibinfo {author} {\bibfnamefont {L.}~\bibnamefont
  {Bernard}}, \bibinfo {author} {\bibfnamefont {L.}~\bibnamefont {Blanchet}},
  \bibinfo {author} {\bibfnamefont {A.}~\bibnamefont {Boh\'e}}, \bibinfo
  {author} {\bibfnamefont {G.}~\bibnamefont {Faye}}, \ and\ \bibinfo {author}
  {\bibfnamefont {S.}~\bibnamefont {Marsat}},\ }\href {\doibase
  10.1103/PhysRevD.93.084037} {\bibfield  {journal} {\bibinfo  {journal} {Phys.
  Rev.}\ }\textbf {\bibinfo {volume} {D93}},\ \bibinfo {pages} {084037}
  (\bibinfo {year} {2016})},\ \Eprint {http://arxiv.org/abs/1512.02876}
  {arXiv:1512.02876 [gr-qc]} \BibitemShut {NoStop}%
\bibitem [{\citenamefont {Marchand}\ \emph {et~al.}(2018)\citenamefont
  {Marchand}, \citenamefont {Bernard}, \citenamefont {Blanchet},\ and\
  \citenamefont {Faye}}]{Marchand:2017pir}%
  \BibitemOpen
  \bibfield  {author} {\bibinfo {author} {\bibfnamefont {T.}~\bibnamefont
  {Marchand}}, \bibinfo {author} {\bibfnamefont {L.}~\bibnamefont {Bernard}},
  \bibinfo {author} {\bibfnamefont {L.}~\bibnamefont {Blanchet}}, \ and\
  \bibinfo {author} {\bibfnamefont {G.}~\bibnamefont {Faye}},\ }\href {\doibase
  10.1103/PhysRevD.97.044023} {\bibfield  {journal} {\bibinfo  {journal} {Phys.
  Rev.}\ }\textbf {\bibinfo {volume} {D97}},\ \bibinfo {pages} {044023}
  (\bibinfo {year} {2018})},\ \Eprint {http://arxiv.org/abs/1707.09289}
  {arXiv:1707.09289 [gr-qc]} \BibitemShut {NoStop}%
\bibitem [{\citenamefont {Foffa}\ and\ \citenamefont
  {Sturani}(2013{\natexlab{a}})}]{Foffa:2012rn}%
  \BibitemOpen
  \bibfield  {author} {\bibinfo {author} {\bibfnamefont {S.}~\bibnamefont
  {Foffa}}\ and\ \bibinfo {author} {\bibfnamefont {R.}~\bibnamefont
  {Sturani}},\ }\href {\doibase 10.1103/PhysRevD.87.064011} {\bibfield
  {journal} {\bibinfo  {journal} {Phys. Rev.}\ }\textbf {\bibinfo {volume}
  {D87}},\ \bibinfo {pages} {064011} (\bibinfo {year} {2013}{\natexlab{a}})},\
  \Eprint {http://arxiv.org/abs/1206.7087} {arXiv:1206.7087 [gr-qc]}
  \BibitemShut {NoStop}%
\bibitem [{\citenamefont {Foffa}\ \emph {et~al.}(2017)\citenamefont {Foffa},
  \citenamefont {Mastrolia}, \citenamefont {Sturani},\ and\ \citenamefont
  {Sturm}}]{Foffa:2016rgu}%
  \BibitemOpen
  \bibfield  {author} {\bibinfo {author} {\bibfnamefont {S.}~\bibnamefont
  {Foffa}}, \bibinfo {author} {\bibfnamefont {P.}~\bibnamefont {Mastrolia}},
  \bibinfo {author} {\bibfnamefont {R.}~\bibnamefont {Sturani}}, \ and\
  \bibinfo {author} {\bibfnamefont {C.}~\bibnamefont {Sturm}},\ }\href
  {\doibase 10.1103/PhysRevD.95.104009} {\bibfield  {journal} {\bibinfo
  {journal} {Phys. Rev.}\ }\textbf {\bibinfo {volume} {D95}},\ \bibinfo {pages}
  {104009} (\bibinfo {year} {2017})},\ \Eprint
  {http://arxiv.org/abs/1612.00482} {arXiv:1612.00482 [gr-qc]} \BibitemShut
  {NoStop}%
\bibitem [{\citenamefont {Foffa}\ and\ \citenamefont
  {Sturani}(2019)}]{Foffa:2019rdf}%
  \BibitemOpen
  \bibfield  {author} {\bibinfo {author} {\bibfnamefont {S.}~\bibnamefont
  {Foffa}}\ and\ \bibinfo {author} {\bibfnamefont {R.}~\bibnamefont
  {Sturani}},\ }\href {\doibase 10.1103/PhysRevD.100.024047} {\bibfield
  {journal} {\bibinfo  {journal} {Phys. Rev.}\ }\textbf {\bibinfo {volume}
  {D100}},\ \bibinfo {pages} {024047} (\bibinfo {year} {2019})},\ \Eprint
  {http://arxiv.org/abs/1903.05113} {arXiv:1903.05113 [gr-qc]} \BibitemShut
  {NoStop}%
\bibitem [{\citenamefont {Porto}\ and\ \citenamefont
  {Rothstein}(2008{\natexlab{a}})}]{Porto:2008tb}%
  \BibitemOpen
  \bibfield  {author} {\bibinfo {author} {\bibfnamefont {R.~A.}\ \bibnamefont
  {Porto}}\ and\ \bibinfo {author} {\bibfnamefont {I.~Z.}\ \bibnamefont
  {Rothstein}},\ }\href {\doibase 10.1103/PhysRevD.78.044012,
  10.1103/PhysRevD.81.029904} {\bibfield  {journal} {\bibinfo  {journal} {Phys.
  Rev.}\ }\textbf {\bibinfo {volume} {D78}},\ \bibinfo {pages} {044012}
  (\bibinfo {year} {2008}{\natexlab{a}})},\ \bibinfo {note} {[Erratum: Phys.
  Rev.D81,029904(2010)]},\ \Eprint {http://arxiv.org/abs/0802.0720}
  {arXiv:0802.0720 [gr-qc]} \BibitemShut {NoStop}%
\bibitem [{\citenamefont {Porto}\ and\ \citenamefont
  {Rothstein}(2008{\natexlab{b}})}]{Porto:2008jj}%
  \BibitemOpen
  \bibfield  {author} {\bibinfo {author} {\bibfnamefont {R.~A.}\ \bibnamefont
  {Porto}}\ and\ \bibinfo {author} {\bibfnamefont {I.~Z.}\ \bibnamefont
  {Rothstein}},\ }\bibfield  {booktitle} {\emph {\bibinfo {booktitle}
  {{Workshop on Effective Field Theory Techniques in Gravitational Wave Physics
  Pittsburgh, Pennsylvania, July 23-25, 2007}}},\ }\href {\doibase
  10.1103/PhysRevD.81.029905, 10.1103/PhysRevD.78.044013} {\bibfield  {journal}
  {\bibinfo  {journal} {Phys. Rev.}\ }\textbf {\bibinfo {volume} {D78}},\
  \bibinfo {pages} {044013} (\bibinfo {year} {2008}{\natexlab{b}})},\ \bibinfo
  {note} {[Erratum: Phys. Rev.D81,029905(2010)]},\ \Eprint
  {http://arxiv.org/abs/0804.0260} {arXiv:0804.0260 [gr-qc]} \BibitemShut
  {NoStop}%
\bibitem [{\citenamefont {Porto}(2010)}]{Porto:2010tr}%
  \BibitemOpen
  \bibfield  {author} {\bibinfo {author} {\bibfnamefont {R.~A.}\ \bibnamefont
  {Porto}},\ }\href {\doibase 10.1088/0264-9381/27/20/205001} {\bibfield
  {journal} {\bibinfo  {journal} {Class. Quant. Grav.}\ }\textbf {\bibinfo
  {volume} {27}},\ \bibinfo {pages} {205001} (\bibinfo {year} {2010})},\
  \Eprint {http://arxiv.org/abs/1005.5730} {arXiv:1005.5730 [gr-qc]}
  \BibitemShut {NoStop}%
\bibitem [{\citenamefont {Porto}\ \emph {et~al.}(2011)\citenamefont {Porto},
  \citenamefont {Ross},\ and\ \citenamefont {Rothstein}}]{Porto:2010zg}%
  \BibitemOpen
  \bibfield  {author} {\bibinfo {author} {\bibfnamefont {R.~A.}\ \bibnamefont
  {Porto}}, \bibinfo {author} {\bibfnamefont {A.}~\bibnamefont {Ross}}, \ and\
  \bibinfo {author} {\bibfnamefont {I.~Z.}\ \bibnamefont {Rothstein}},\ }\href
  {\doibase 10.1088/1475-7516/2011/03/009} {\bibfield  {journal} {\bibinfo
  {journal} {JCAP}\ }\textbf {\bibinfo {volume} {1103}},\ \bibinfo {pages}
  {009} (\bibinfo {year} {2011})},\ \Eprint {http://arxiv.org/abs/1007.1312}
  {arXiv:1007.1312 [gr-qc]} \BibitemShut {NoStop}%
\bibitem [{\citenamefont {Levi}(2012)}]{Levi:2011eq}%
  \BibitemOpen
  \bibfield  {author} {\bibinfo {author} {\bibfnamefont {M.}~\bibnamefont
  {Levi}},\ }\href {\doibase 10.1103/PhysRevD.85.064043} {\bibfield  {journal}
  {\bibinfo  {journal} {Phys. Rev.}\ }\textbf {\bibinfo {volume} {D85}},\
  \bibinfo {pages} {064043} (\bibinfo {year} {2012})},\ \Eprint
  {http://arxiv.org/abs/1107.4322} {arXiv:1107.4322 [gr-qc]} \BibitemShut
  {NoStop}%
\bibitem [{\citenamefont {Porto}\ \emph {et~al.}(2012)\citenamefont {Porto},
  \citenamefont {Ross},\ and\ \citenamefont {Rothstein}}]{Porto:2012as}%
  \BibitemOpen
  \bibfield  {author} {\bibinfo {author} {\bibfnamefont {R.~A.}\ \bibnamefont
  {Porto}}, \bibinfo {author} {\bibfnamefont {A.}~\bibnamefont {Ross}}, \ and\
  \bibinfo {author} {\bibfnamefont {I.~Z.}\ \bibnamefont {Rothstein}},\ }\href
  {\doibase 10.1088/1475-7516/2012/09/028} {\bibfield  {journal} {\bibinfo
  {journal} {JCAP}\ }\textbf {\bibinfo {volume} {1209}},\ \bibinfo {pages}
  {028} (\bibinfo {year} {2012})},\ \Eprint {http://arxiv.org/abs/1203.2962}
  {arXiv:1203.2962 [gr-qc]} \BibitemShut {NoStop}%
\bibitem [{\citenamefont {Levi}\ and\ \citenamefont
  {Steinhoff}(2016{\natexlab{a}})}]{Levi:2015uxa}%
  \BibitemOpen
  \bibfield  {author} {\bibinfo {author} {\bibfnamefont {M.}~\bibnamefont
  {Levi}}\ and\ \bibinfo {author} {\bibfnamefont {J.}~\bibnamefont
  {Steinhoff}},\ }\href {\doibase 10.1088/1475-7516/2016/01/011} {\bibfield
  {journal} {\bibinfo  {journal} {JCAP}\ }\textbf {\bibinfo {volume} {1601}},\
  \bibinfo {pages} {011} (\bibinfo {year} {2016}{\natexlab{a}})},\ \Eprint
  {http://arxiv.org/abs/1506.05056} {arXiv:1506.05056 [gr-qc]} \BibitemShut
  {NoStop}%
\bibitem [{\citenamefont {Levi}\ and\ \citenamefont
  {Steinhoff}(2016{\natexlab{b}})}]{Levi:2015ixa}%
  \BibitemOpen
  \bibfield  {author} {\bibinfo {author} {\bibfnamefont {M.}~\bibnamefont
  {Levi}}\ and\ \bibinfo {author} {\bibfnamefont {J.}~\bibnamefont
  {Steinhoff}},\ }\href {\doibase 10.1088/1475-7516/2016/01/008} {\bibfield
  {journal} {\bibinfo  {journal} {JCAP}\ }\textbf {\bibinfo {volume} {1601}},\
  \bibinfo {pages} {008} (\bibinfo {year} {2016}{\natexlab{b}})},\ \Eprint
  {http://arxiv.org/abs/1506.05794} {arXiv:1506.05794 [gr-qc]} \BibitemShut
  {NoStop}%
\bibitem [{\citenamefont {Levi}\ and\ \citenamefont
  {Steinhoff}(2015)}]{Levi:2014gsa}%
  \BibitemOpen
  \bibfield  {author} {\bibinfo {author} {\bibfnamefont {M.}~\bibnamefont
  {Levi}}\ and\ \bibinfo {author} {\bibfnamefont {J.}~\bibnamefont
  {Steinhoff}},\ }\href {\doibase 10.1007/JHEP06(2015)059} {\bibfield
  {journal} {\bibinfo  {journal} {JHEP}\ }\textbf {\bibinfo {volume} {06}},\
  \bibinfo {pages} {059} (\bibinfo {year} {2015})},\ \Eprint
  {http://arxiv.org/abs/1410.2601} {arXiv:1410.2601 [gr-qc]} \BibitemShut
  {NoStop}%
\bibitem [{\citenamefont {Levi}\ and\ \citenamefont
  {Steinhoff}(2016{\natexlab{c}})}]{Levi:2016ofk}%
  \BibitemOpen
  \bibfield  {author} {\bibinfo {author} {\bibfnamefont {M.}~\bibnamefont
  {Levi}}\ and\ \bibinfo {author} {\bibfnamefont {J.}~\bibnamefont
  {Steinhoff}},\ }\href@noop {} {\  (\bibinfo {year} {2016}{\natexlab{c}})},\
  \Eprint {http://arxiv.org/abs/1607.04252} {arXiv:1607.04252 [gr-qc]}
  \BibitemShut {NoStop}%
\bibitem [{\citenamefont {Blanchet}\ \emph {et~al.}(2002)\citenamefont
  {Blanchet}, \citenamefont {Faye}, \citenamefont {Iyer},\ and\ \citenamefont
  {Joguet}}]{Blanchet:2001ax}%
  \BibitemOpen
  \bibfield  {author} {\bibinfo {author} {\bibfnamefont {L.}~\bibnamefont
  {Blanchet}}, \bibinfo {author} {\bibfnamefont {G.}~\bibnamefont {Faye}},
  \bibinfo {author} {\bibfnamefont {B.~R.}\ \bibnamefont {Iyer}}, \ and\
  \bibinfo {author} {\bibfnamefont {B.}~\bibnamefont {Joguet}},\ }\href
  {\doibase 10.1103/PhysRevD.71.129902, 10.1103/PhysRevD.65.061501} {\bibfield
  {journal} {\bibinfo  {journal} {Phys. Rev.}\ }\textbf {\bibinfo {volume}
  {D65}},\ \bibinfo {pages} {061501} (\bibinfo {year} {2002})},\ \bibinfo
  {note} {[Erratum: Phys. Rev.D71,129902(2005)]},\ \Eprint
  {http://arxiv.org/abs/gr-qc/0105099} {arXiv:gr-qc/0105099 [gr-qc]}
  \BibitemShut {NoStop}%
\bibitem [{\citenamefont {Blanchet}\ \emph {et~al.}(2004)\citenamefont
  {Blanchet}, \citenamefont {Damour}, \citenamefont {Esposito-Farese},\ and\
  \citenamefont {Iyer}}]{Blanchet:2004ek}%
  \BibitemOpen
  \bibfield  {author} {\bibinfo {author} {\bibfnamefont {L.}~\bibnamefont
  {Blanchet}}, \bibinfo {author} {\bibfnamefont {T.}~\bibnamefont {Damour}},
  \bibinfo {author} {\bibfnamefont {G.}~\bibnamefont {Esposito-Farese}}, \ and\
  \bibinfo {author} {\bibfnamefont {B.~R.}\ \bibnamefont {Iyer}},\ }\href
  {\doibase 10.1103/PhysRevLett.93.091101} {\bibfield  {journal} {\bibinfo
  {journal} {Phys. Rev. Lett.}\ }\textbf {\bibinfo {volume} {93}},\ \bibinfo
  {pages} {091101} (\bibinfo {year} {2004})},\ \Eprint
  {http://arxiv.org/abs/gr-qc/0406012} {arXiv:gr-qc/0406012 [gr-qc]}
  \BibitemShut {NoStop}%
\bibitem [{\citenamefont {Boh\'e}\ \emph {et~al.}(2013)\citenamefont {Boh\'e},
  \citenamefont {Marsat},\ and\ \citenamefont {Blanchet}}]{Bohe:2013cla}%
  \BibitemOpen
  \bibfield  {author} {\bibinfo {author} {\bibfnamefont {A.}~\bibnamefont
  {Boh\'e}}, \bibinfo {author} {\bibfnamefont {S.}~\bibnamefont {Marsat}}, \
  and\ \bibinfo {author} {\bibfnamefont {L.}~\bibnamefont {Blanchet}},\ }\href
  {\doibase 10.1088/0264-9381/30/13/135009} {\bibfield  {journal} {\bibinfo
  {journal} {Class. Quant. Grav.}\ }\textbf {\bibinfo {volume} {30}},\ \bibinfo
  {pages} {135009} (\bibinfo {year} {2013})},\ \Eprint
  {http://arxiv.org/abs/1303.7412} {arXiv:1303.7412 [gr-qc]} \BibitemShut
  {NoStop}%
\bibitem [{\citenamefont {Marsat}\ \emph {et~al.}(2014)\citenamefont {Marsat},
  \citenamefont {Boh\'e}, \citenamefont {Blanchet},\ and\ \citenamefont
  {Buonanno}}]{Marsat:2013caa}%
  \BibitemOpen
  \bibfield  {author} {\bibinfo {author} {\bibfnamefont {S.}~\bibnamefont
  {Marsat}}, \bibinfo {author} {\bibfnamefont {A.}~\bibnamefont {Boh\'e}},
  \bibinfo {author} {\bibfnamefont {L.}~\bibnamefont {Blanchet}}, \ and\
  \bibinfo {author} {\bibfnamefont {A.}~\bibnamefont {Buonanno}},\ }\href
  {\doibase 10.1088/0264-9381/31/2/025023} {\bibfield  {journal} {\bibinfo
  {journal} {Class. Quant. Grav.}\ }\textbf {\bibinfo {volume} {31}},\ \bibinfo
  {pages} {025023} (\bibinfo {year} {2014})},\ \Eprint
  {http://arxiv.org/abs/1307.6793} {arXiv:1307.6793 [gr-qc]} \BibitemShut
  {NoStop}%
\bibitem [{\citenamefont {Boh{\'e}}\ \emph {et~al.}(2015)\citenamefont
  {Boh{\'e}}, \citenamefont {Faye}, \citenamefont {Marsat},\ and\ \citenamefont
  {Porter}}]{Bohe:2015ana}%
  \BibitemOpen
  \bibfield  {author} {\bibinfo {author} {\bibfnamefont {A.}~\bibnamefont
  {Boh{\'e}}}, \bibinfo {author} {\bibfnamefont {G.}~\bibnamefont {Faye}},
  \bibinfo {author} {\bibfnamefont {S.}~\bibnamefont {Marsat}}, \ and\ \bibinfo
  {author} {\bibfnamefont {E.~K.}\ \bibnamefont {Porter}},\ }\href {\doibase
  10.1088/0264-9381/32/19/195010} {\bibfield  {journal} {\bibinfo  {journal}
  {Class. Quant. Grav.}\ }\textbf {\bibinfo {volume} {32}},\ \bibinfo {pages}
  {195010} (\bibinfo {year} {2015})},\ \Eprint
  {http://arxiv.org/abs/1501.01529} {arXiv:1501.01529 [gr-qc]} \BibitemShut
  {NoStop}%
\bibitem [{\citenamefont {Marsat}(2015)}]{Marsat:2014xea}%
  \BibitemOpen
  \bibfield  {author} {\bibinfo {author} {\bibfnamefont {S.}~\bibnamefont
  {Marsat}},\ }\href {\doibase 10.1088/0264-9381/32/8/085008} {\bibfield
  {journal} {\bibinfo  {journal} {Class. Quant. Grav.}\ }\textbf {\bibinfo
  {volume} {32}},\ \bibinfo {pages} {085008} (\bibinfo {year} {2015})},\
  \Eprint {http://arxiv.org/abs/1411.4118} {arXiv:1411.4118 [gr-qc]}
  \BibitemShut {NoStop}%
\bibitem [{\citenamefont {Siemonsen}\ \emph {et~al.}(2018)\citenamefont
  {Siemonsen}, \citenamefont {Steinhoff},\ and\ \citenamefont
  {Vines}}]{Siemonsen:2017yux}%
  \BibitemOpen
  \bibfield  {author} {\bibinfo {author} {\bibfnamefont {N.}~\bibnamefont
  {Siemonsen}}, \bibinfo {author} {\bibfnamefont {J.}~\bibnamefont
  {Steinhoff}}, \ and\ \bibinfo {author} {\bibfnamefont {J.}~\bibnamefont
  {Vines}},\ }\href {\doibase 10.1103/PhysRevD.97.124046} {\bibfield  {journal}
  {\bibinfo  {journal} {Phys. Rev.}\ }\textbf {\bibinfo {volume} {D97}},\
  \bibinfo {pages} {124046} (\bibinfo {year} {2018})},\ \Eprint
  {http://arxiv.org/abs/1712.08603} {arXiv:1712.08603 [gr-qc]} \BibitemShut
  {NoStop}%
\bibitem [{\citenamefont {Damour}(1982)}]{Damour:1982wm}%
  \BibitemOpen
  \bibfield  {author} {\bibinfo {author} {\bibfnamefont {T.}~\bibnamefont
  {Damour}},\ }in\ \href@noop {} {\emph {\bibinfo {booktitle} {{Les Houches
  Summer School on Gravitational Radiation Les Houches, France, June 2-21,
  1982}}}}\ (\bibinfo {year} {1982})\BibitemShut {NoStop}%
\bibitem [{\citenamefont {Hinderer}(2008)}]{Hinderer:2007mb}%
  \BibitemOpen
  \bibfield  {author} {\bibinfo {author} {\bibfnamefont {T.}~\bibnamefont
  {Hinderer}},\ }\href {\doibase 10.1086/533487} {\bibfield  {journal}
  {\bibinfo  {journal} {Astrophys. J.}\ }\textbf {\bibinfo {volume} {677}},\
  \bibinfo {pages} {1216} (\bibinfo {year} {2008})},\ \Eprint
  {http://arxiv.org/abs/0711.2420} {arXiv:0711.2420 [astro-ph]} \BibitemShut
  {NoStop}%
\bibitem [{\citenamefont {Abbott}\ \emph {et~al.}(2017)\citenamefont {Abbott}
  \emph {et~al.}}]{TheLIGOScientific:2017qsa}%
  \BibitemOpen
  \bibfield  {author} {\bibinfo {author} {\bibfnamefont {B.}~\bibnamefont
  {Abbott}} \emph {et~al.} (\bibinfo {collaboration} {Virgo, LIGO
  Scientific}),\ }\href {\doibase 10.1103/PhysRevLett.119.161101} {\bibfield
  {journal} {\bibinfo  {journal} {Phys. Rev. Lett.}\ }\textbf {\bibinfo
  {volume} {119}},\ \bibinfo {pages} {161101} (\bibinfo {year} {2017})},\
  \Eprint {http://arxiv.org/abs/1710.05832} {arXiv:1710.05832 [gr-qc]}
  \BibitemShut {NoStop}%
\bibitem [{\citenamefont {Abbott}\ \emph
  {et~al.}(2019{\natexlab{b}})\citenamefont {Abbott} \emph
  {et~al.}}]{Abbott:2018wiz}%
  \BibitemOpen
  \bibfield  {author} {\bibinfo {author} {\bibfnamefont {B.~P.}\ \bibnamefont
  {Abbott}} \emph {et~al.} (\bibinfo {collaboration} {LIGO Scientific,
  Virgo}),\ }\href {\doibase 10.1103/PhysRevX.9.011001} {\bibfield  {journal}
  {\bibinfo  {journal} {Phys. Rev.}\ }\textbf {\bibinfo {volume} {X9}},\
  \bibinfo {pages} {011001} (\bibinfo {year} {2019}{\natexlab{b}})},\ \Eprint
  {http://arxiv.org/abs/1805.11579} {arXiv:1805.11579 [gr-qc]} \BibitemShut
  {NoStop}%
\bibitem [{\citenamefont {Binnington}\ and\ \citenamefont
  {Poisson}(2009)}]{Binnington:2009bb}%
  \BibitemOpen
  \bibfield  {author} {\bibinfo {author} {\bibfnamefont {T.}~\bibnamefont
  {Binnington}}\ and\ \bibinfo {author} {\bibfnamefont {E.}~\bibnamefont
  {Poisson}},\ }\href {\doibase 10.1103/PhysRevD.80.084018} {\bibfield
  {journal} {\bibinfo  {journal} {Phys. Rev.}\ }\textbf {\bibinfo {volume}
  {D80}},\ \bibinfo {pages} {084018} (\bibinfo {year} {2009})},\ \Eprint
  {http://arxiv.org/abs/0906.1366} {arXiv:0906.1366 [gr-qc]} \BibitemShut
  {NoStop}%
\bibitem [{\citenamefont {Damour}\ and\ \citenamefont
  {Nagar}(2009)}]{Damour:2009vw}%
  \BibitemOpen
  \bibfield  {author} {\bibinfo {author} {\bibfnamefont {T.}~\bibnamefont
  {Damour}}\ and\ \bibinfo {author} {\bibfnamefont {A.}~\bibnamefont {Nagar}},\
  }\href {\doibase 10.1103/PhysRevD.80.084035} {\bibfield  {journal} {\bibinfo
  {journal} {Phys. Rev.}\ }\textbf {\bibinfo {volume} {D80}},\ \bibinfo {pages}
  {084035} (\bibinfo {year} {2009})},\ \Eprint {http://arxiv.org/abs/0906.0096}
  {arXiv:0906.0096 [gr-qc]} \BibitemShut {NoStop}%
\bibitem [{\citenamefont {Kol}\ and\ \citenamefont
  {Smolkin}(2012)}]{Kol:2011vg}%
  \BibitemOpen
  \bibfield  {author} {\bibinfo {author} {\bibfnamefont {B.}~\bibnamefont
  {Kol}}\ and\ \bibinfo {author} {\bibfnamefont {M.}~\bibnamefont {Smolkin}},\
  }\href {\doibase 10.1007/JHEP02(2012)010} {\bibfield  {journal} {\bibinfo
  {journal} {JHEP}\ }\textbf {\bibinfo {volume} {02}},\ \bibinfo {pages} {010}
  (\bibinfo {year} {2012})},\ \Eprint {http://arxiv.org/abs/1110.3764}
  {arXiv:1110.3764 [hep-th]} \BibitemShut {NoStop}%
\bibitem [{\citenamefont {Gürlebeck}(2015)}]{Gurlebeck:2015xpa}%
  \BibitemOpen
  \bibfield  {author} {\bibinfo {author} {\bibfnamefont {N.}~\bibnamefont
  {Gürlebeck}},\ }\href {\doibase 10.1103/PhysRevLett.114.151102} {\bibfield
  {journal} {\bibinfo  {journal} {Phys. Rev. Lett.}\ }\textbf {\bibinfo
  {volume} {114}},\ \bibinfo {pages} {151102} (\bibinfo {year} {2015})},\
  \Eprint {http://arxiv.org/abs/1503.03240} {arXiv:1503.03240 [gr-qc]}
  \BibitemShut {NoStop}%
\bibitem [{\citenamefont {Pani}\ \emph
  {et~al.}(2015{\natexlab{a}})\citenamefont {Pani}, \citenamefont {Gualtieri},
  \citenamefont {Maselli},\ and\ \citenamefont {Ferrari}}]{Pani:2015hfa}%
  \BibitemOpen
  \bibfield  {author} {\bibinfo {author} {\bibfnamefont {P.}~\bibnamefont
  {Pani}}, \bibinfo {author} {\bibfnamefont {L.}~\bibnamefont {Gualtieri}},
  \bibinfo {author} {\bibfnamefont {A.}~\bibnamefont {Maselli}}, \ and\
  \bibinfo {author} {\bibfnamefont {V.}~\bibnamefont {Ferrari}},\ }\href
  {\doibase 10.1103/PhysRevD.92.024010} {\bibfield  {journal} {\bibinfo
  {journal} {Phys. Rev.}\ }\textbf {\bibinfo {volume} {D92}},\ \bibinfo {pages}
  {024010} (\bibinfo {year} {2015}{\natexlab{a}})},\ \Eprint
  {http://arxiv.org/abs/1503.07365} {arXiv:1503.07365 [gr-qc]} \BibitemShut
  {NoStop}%
\bibitem [{\citenamefont {Pani}\ \emph
  {et~al.}(2015{\natexlab{b}})\citenamefont {Pani}, \citenamefont {Gualtieri},\
  and\ \citenamefont {Ferrari}}]{Pani:2015nua}%
  \BibitemOpen
  \bibfield  {author} {\bibinfo {author} {\bibfnamefont {P.}~\bibnamefont
  {Pani}}, \bibinfo {author} {\bibfnamefont {L.}~\bibnamefont {Gualtieri}}, \
  and\ \bibinfo {author} {\bibfnamefont {V.}~\bibnamefont {Ferrari}},\ }\href
  {\doibase 10.1103/PhysRevD.92.124003} {\bibfield  {journal} {\bibinfo
  {journal} {Phys. Rev.}\ }\textbf {\bibinfo {volume} {D92}},\ \bibinfo {pages}
  {124003} (\bibinfo {year} {2015}{\natexlab{b}})},\ \Eprint
  {http://arxiv.org/abs/1509.02171} {arXiv:1509.02171 [gr-qc]} \BibitemShut
  {NoStop}%
\bibitem [{\citenamefont {Landry}\ and\ \citenamefont
  {Poisson}(2015)}]{Landry:2015zfa}%
  \BibitemOpen
  \bibfield  {author} {\bibinfo {author} {\bibfnamefont {P.}~\bibnamefont
  {Landry}}\ and\ \bibinfo {author} {\bibfnamefont {E.}~\bibnamefont
  {Poisson}},\ }\href {\doibase 10.1103/PhysRevD.91.104018} {\bibfield
  {journal} {\bibinfo  {journal} {Phys. Rev.}\ }\textbf {\bibinfo {volume}
  {D91}},\ \bibinfo {pages} {104018} (\bibinfo {year} {2015})},\ \Eprint
  {http://arxiv.org/abs/1503.07366} {arXiv:1503.07366 [gr-qc]} \BibitemShut
  {NoStop}%
\bibitem [{\citenamefont {Foffa}\ \emph
  {et~al.}(2019{\natexlab{a}})\citenamefont {Foffa}, \citenamefont {Porto},
  \citenamefont {Rothstein},\ and\ \citenamefont {Sturani}}]{Foffa:2019yfl}%
  \BibitemOpen
  \bibfield  {author} {\bibinfo {author} {\bibfnamefont {S.}~\bibnamefont
  {Foffa}}, \bibinfo {author} {\bibfnamefont {R.~A.}\ \bibnamefont {Porto}},
  \bibinfo {author} {\bibfnamefont {I.}~\bibnamefont {Rothstein}}, \ and\
  \bibinfo {author} {\bibfnamefont {R.}~\bibnamefont {Sturani}},\ }\href
  {\doibase 10.1103/PhysRevD.100.024048} {\bibfield  {journal} {\bibinfo
  {journal} {Phys. Rev.}\ }\textbf {\bibinfo {volume} {D100}},\ \bibinfo
  {pages} {024048} (\bibinfo {year} {2019}{\natexlab{a}})},\ \Eprint
  {http://arxiv.org/abs/1903.05118} {arXiv:1903.05118 [gr-qc]} \BibitemShut
  {NoStop}%
\bibitem [{\citenamefont {Blanchet}\ and\ \citenamefont
  {Damour}(1988)}]{Blanchet:1987wq}%
  \BibitemOpen
  \bibfield  {author} {\bibinfo {author} {\bibfnamefont {L.}~\bibnamefont
  {Blanchet}}\ and\ \bibinfo {author} {\bibfnamefont {T.}~\bibnamefont
  {Damour}},\ }\href {\doibase 10.1103/PhysRevD.37.1410} {\bibfield  {journal}
  {\bibinfo  {journal} {Phys. Rev.}\ }\textbf {\bibinfo {volume} {D37}},\
  \bibinfo {pages} {1410} (\bibinfo {year} {1988})}\BibitemShut {NoStop}%
\bibitem [{\citenamefont {Blanchet}\ and\ \citenamefont
  {Damour}(1992)}]{Blanchet:1992br}%
  \BibitemOpen
  \bibfield  {author} {\bibinfo {author} {\bibfnamefont {L.}~\bibnamefont
  {Blanchet}}\ and\ \bibinfo {author} {\bibfnamefont {T.}~\bibnamefont
  {Damour}},\ }\href {\doibase 10.1103/PhysRevD.46.4304} {\bibfield  {journal}
  {\bibinfo  {journal} {Phys. Rev.}\ }\textbf {\bibinfo {volume} {D46}},\
  \bibinfo {pages} {4304} (\bibinfo {year} {1992})}\BibitemShut {NoStop}%
\bibitem [{\citenamefont {Blanchet}\ and\ \citenamefont
  {Sch{\"a}fer}(1993)}]{Blanchet:1993ec}%
  \BibitemOpen
  \bibfield  {author} {\bibinfo {author} {\bibfnamefont {L.}~\bibnamefont
  {Blanchet}}\ and\ \bibinfo {author} {\bibfnamefont {G.}~\bibnamefont
  {Sch{\"a}fer}},\ }\href {\doibase 10.1088/0264-9381/10/12/026} {\bibfield
  {journal} {\bibinfo  {journal} {Class. Quant. Grav.}\ }\textbf {\bibinfo
  {volume} {10}},\ \bibinfo {pages} {2699} (\bibinfo {year}
  {1993})}\BibitemShut {NoStop}%
\bibitem [{\citenamefont {Goldberger}\ and\ \citenamefont
  {Ross}(2010)}]{Goldberger:2009qd}%
  \BibitemOpen
  \bibfield  {author} {\bibinfo {author} {\bibfnamefont {W.~D.}\ \bibnamefont
  {Goldberger}}\ and\ \bibinfo {author} {\bibfnamefont {A.}~\bibnamefont
  {Ross}},\ }\href {\doibase 10.1103/PhysRevD.81.124015} {\bibfield  {journal}
  {\bibinfo  {journal} {Phys. Rev.}\ }\textbf {\bibinfo {volume} {D81}},\
  \bibinfo {pages} {124015} (\bibinfo {year} {2010})},\ \Eprint
  {http://arxiv.org/abs/0912.4254} {arXiv:0912.4254 [gr-qc]} \BibitemShut
  {NoStop}%
\bibitem [{\citenamefont {Christodoulou}(1991)}]{Christodoulou:1991cr}%
  \BibitemOpen
  \bibfield  {author} {\bibinfo {author} {\bibfnamefont {D.}~\bibnamefont
  {Christodoulou}},\ }\href {\doibase 10.1103/PhysRevLett.67.1486} {\bibfield
  {journal} {\bibinfo  {journal} {Phys. Rev. Lett.}\ }\textbf {\bibinfo
  {volume} {67}},\ \bibinfo {pages} {1486} (\bibinfo {year}
  {1991})}\BibitemShut {NoStop}%
\bibitem [{\citenamefont {{Zel'dovich}}\ and\ \citenamefont
  {{Polnarev}}(1974)}]{1974SvA....18...17Z}%
  \BibitemOpen
  \bibfield  {author} {\bibinfo {author} {\bibfnamefont {Y.~B.}\ \bibnamefont
  {{Zel'dovich}}}\ and\ \bibinfo {author} {\bibfnamefont {A.~G.}\ \bibnamefont
  {{Polnarev}}},\ }\href@noop {} {\bibfield  {journal} {\bibinfo  {journal}
  {Sov. Ast.}\ }\textbf {\bibinfo {volume} {18}},\ \bibinfo {pages} {17}
  (\bibinfo {year} {1974})}\BibitemShut {NoStop}%
\bibitem [{\citenamefont {Thorne}(1992)}]{Thorne:1992sdb}%
  \BibitemOpen
  \bibfield  {author} {\bibinfo {author} {\bibfnamefont {K.~S.}\ \bibnamefont
  {Thorne}},\ }\href {\doibase 10.1103/PhysRevD.45.520} {\bibfield  {journal}
  {\bibinfo  {journal} {Phys. Rev.}\ }\textbf {\bibinfo {volume} {D45}},\
  \bibinfo {pages} {520} (\bibinfo {year} {1992})}\BibitemShut {NoStop}%
\bibitem [{\citenamefont {Wiseman}\ and\ \citenamefont
  {Will}(1991)}]{Wiseman:1991ss}%
  \BibitemOpen
  \bibfield  {author} {\bibinfo {author} {\bibfnamefont {A.~G.}\ \bibnamefont
  {Wiseman}}\ and\ \bibinfo {author} {\bibfnamefont {C.~M.}\ \bibnamefont
  {Will}},\ }\href {\doibase 10.1103/PhysRevD.44.R2945} {\bibfield  {journal}
  {\bibinfo  {journal} {Phys. Rev.}\ }\textbf {\bibinfo {volume} {D44}},\
  \bibinfo {pages} {R2945} (\bibinfo {year} {1991})}\BibitemShut {NoStop}%
\bibitem [{\citenamefont {Foffa}\ and\ \citenamefont
  {Sturani}(2013{\natexlab{b}})}]{Foffa:2011np}%
  \BibitemOpen
  \bibfield  {author} {\bibinfo {author} {\bibfnamefont {S.}~\bibnamefont
  {Foffa}}\ and\ \bibinfo {author} {\bibfnamefont {R.}~\bibnamefont
  {Sturani}},\ }\href {\doibase 10.1103/PhysRevD.87.044056} {\bibfield
  {journal} {\bibinfo  {journal} {Phys. Rev.}\ }\textbf {\bibinfo {volume}
  {D87}},\ \bibinfo {pages} {044056} (\bibinfo {year} {2013}{\natexlab{b}})},\
  \Eprint {http://arxiv.org/abs/1111.5488} {arXiv:1111.5488 [gr-qc]}
  \BibitemShut {NoStop}%
\bibitem [{\citenamefont {Galley}\ \emph {et~al.}(2016)\citenamefont {Galley},
  \citenamefont {Leibovich}, \citenamefont {Porto},\ and\ \citenamefont
  {Ross}}]{Galley:2015kus}%
  \BibitemOpen
  \bibfield  {author} {\bibinfo {author} {\bibfnamefont {C.~R.}\ \bibnamefont
  {Galley}}, \bibinfo {author} {\bibfnamefont {A.~K.}\ \bibnamefont
  {Leibovich}}, \bibinfo {author} {\bibfnamefont {R.~A.}\ \bibnamefont
  {Porto}}, \ and\ \bibinfo {author} {\bibfnamefont {A.}~\bibnamefont {Ross}},\
  }\href {\doibase 10.1103/PhysRevD.93.124010} {\bibfield  {journal} {\bibinfo
  {journal} {Phys. Rev.}\ }\textbf {\bibinfo {volume} {D93}},\ \bibinfo {pages}
  {124010} (\bibinfo {year} {2016})},\ \Eprint
  {http://arxiv.org/abs/1511.07379} {arXiv:1511.07379 [gr-qc]} \BibitemShut
  {NoStop}%
\bibitem [{\citenamefont {Manohar}\ and\ \citenamefont
  {Stewart}(2007)}]{Manohar:2006nz}%
  \BibitemOpen
  \bibfield  {author} {\bibinfo {author} {\bibfnamefont {A.~V.}\ \bibnamefont
  {Manohar}}\ and\ \bibinfo {author} {\bibfnamefont {I.~W.}\ \bibnamefont
  {Stewart}},\ }\href {\doibase 10.1103/PhysRevD.76.074002} {\bibfield
  {journal} {\bibinfo  {journal} {Phys. Rev.}\ }\textbf {\bibinfo {volume}
  {D76}},\ \bibinfo {pages} {074002} (\bibinfo {year} {2007})},\ \Eprint
  {http://arxiv.org/abs/hep-ph/0605001} {arXiv:hep-ph/0605001 [hep-ph]}
  \BibitemShut {NoStop}%
\bibitem [{\citenamefont {Le~Tiec}\ \emph {et~al.}(2012)\citenamefont
  {Le~Tiec}, \citenamefont {Blanchet},\ and\ \citenamefont
  {Whiting}}]{LeTiec:2011ab}%
  \BibitemOpen
  \bibfield  {author} {\bibinfo {author} {\bibfnamefont {A.}~\bibnamefont
  {Le~Tiec}}, \bibinfo {author} {\bibfnamefont {L.}~\bibnamefont {Blanchet}}, \
  and\ \bibinfo {author} {\bibfnamefont {B.~F.}\ \bibnamefont {Whiting}},\
  }\href {\doibase 10.1103/PhysRevD.85.064039} {\bibfield  {journal} {\bibinfo
  {journal} {Phys. Rev.}\ }\textbf {\bibinfo {volume} {D85}},\ \bibinfo {pages}
  {064039} (\bibinfo {year} {2012})},\ \Eprint {http://arxiv.org/abs/1111.5378}
  {arXiv:1111.5378 [gr-qc]} \BibitemShut {NoStop}%
\bibitem [{\citenamefont {Bini}\ and\ \citenamefont
  {Damour}(2014)}]{Bini:2013rfa}%
  \BibitemOpen
  \bibfield  {author} {\bibinfo {author} {\bibfnamefont {D.}~\bibnamefont
  {Bini}}\ and\ \bibinfo {author} {\bibfnamefont {T.}~\bibnamefont {Damour}},\
  }\href {\doibase 10.1103/PhysRevD.89.064063} {\bibfield  {journal} {\bibinfo
  {journal} {Phys. Rev.}\ }\textbf {\bibinfo {volume} {D89}},\ \bibinfo {pages}
  {064063} (\bibinfo {year} {2014})},\ \Eprint {http://arxiv.org/abs/1312.2503}
  {arXiv:1312.2503 [gr-qc]} \BibitemShut {NoStop}%
\bibitem [{\citenamefont {Barack}\ \emph {et~al.}(2010)\citenamefont {Barack},
  \citenamefont {Damour},\ and\ \citenamefont {Sago}}]{Barack:2010ny}%
  \BibitemOpen
  \bibfield  {author} {\bibinfo {author} {\bibfnamefont {L.}~\bibnamefont
  {Barack}}, \bibinfo {author} {\bibfnamefont {T.}~\bibnamefont {Damour}}, \
  and\ \bibinfo {author} {\bibfnamefont {N.}~\bibnamefont {Sago}},\ }\href
  {\doibase 10.1103/PhysRevD.82.084036} {\bibfield  {journal} {\bibinfo
  {journal} {Phys. Rev.}\ }\textbf {\bibinfo {volume} {D82}},\ \bibinfo {pages}
  {084036} (\bibinfo {year} {2010})},\ \Eprint {http://arxiv.org/abs/1008.0935}
  {arXiv:1008.0935 [gr-qc]} \BibitemShut {NoStop}%
\bibitem [{\citenamefont {Porto}(2017)}]{Porto:2017shd}%
  \BibitemOpen
  \bibfield  {author} {\bibinfo {author} {\bibfnamefont {R.~A.}\ \bibnamefont
  {Porto}},\ }\href {\doibase 10.1103/PhysRevD.96.024063} {\bibfield  {journal}
  {\bibinfo  {journal} {Phys. Rev.}\ }\textbf {\bibinfo {volume} {D96}},\
  \bibinfo {pages} {024063} (\bibinfo {year} {2017})},\ \Eprint
  {http://arxiv.org/abs/1703.06434} {arXiv:1703.06434 [gr-qc]} \BibitemShut
  {NoStop}%
\bibitem [{\citenamefont {Porto}\ and\ \citenamefont
  {Rothstein}(2017)}]{Porto:2017dgs}%
  \BibitemOpen
  \bibfield  {author} {\bibinfo {author} {\bibfnamefont {R.~A.}\ \bibnamefont
  {Porto}}\ and\ \bibinfo {author} {\bibfnamefont {I.~Z.}\ \bibnamefont
  {Rothstein}},\ }\href {\doibase 10.1103/PhysRevD.96.024062} {\bibfield
  {journal} {\bibinfo  {journal} {Phys. Rev.}\ }\textbf {\bibinfo {volume}
  {D96}},\ \bibinfo {pages} {024062} (\bibinfo {year} {2017})},\ \Eprint
  {http://arxiv.org/abs/1703.06433} {arXiv:1703.06433 [gr-qc]} \BibitemShut
  {NoStop}%
\bibitem [{\citenamefont {Kol}\ and\ \citenamefont
  {Smolkin}(2008)}]{Kol:2007bc}%
  \BibitemOpen
  \bibfield  {author} {\bibinfo {author} {\bibfnamefont {B.}~\bibnamefont
  {Kol}}\ and\ \bibinfo {author} {\bibfnamefont {M.}~\bibnamefont {Smolkin}},\
  }\href {\doibase 10.1088/0264-9381/25/14/145011} {\bibfield  {journal}
  {\bibinfo  {journal} {Class. Quant. Grav.}\ }\textbf {\bibinfo {volume}
  {25}},\ \bibinfo {pages} {145011} (\bibinfo {year} {2008})},\ \Eprint
  {http://arxiv.org/abs/0712.4116} {arXiv:0712.4116 [hep-th]} \BibitemShut
  {NoStop}%
\bibitem [{\citenamefont {Galley}\ and\ \citenamefont
  {Tiglio}(2009)}]{Galley:2009px}%
  \BibitemOpen
  \bibfield  {author} {\bibinfo {author} {\bibfnamefont {C.~R.}\ \bibnamefont
  {Galley}}\ and\ \bibinfo {author} {\bibfnamefont {M.}~\bibnamefont
  {Tiglio}},\ }\href {\doibase 10.1103/PhysRevD.79.124027} {\bibfield
  {journal} {\bibinfo  {journal} {Phys. Rev.}\ }\textbf {\bibinfo {volume}
  {D79}},\ \bibinfo {pages} {124027} (\bibinfo {year} {2009})},\ \Eprint
  {http://arxiv.org/abs/0903.1122} {arXiv:0903.1122 [gr-qc]} \BibitemShut
  {NoStop}%
\bibitem [{\citenamefont {Foffa}\ and\ \citenamefont {Sturani}(2021)}]{FS21}%
  \BibitemOpen
  \bibfield  {author} {\bibinfo {author} {\bibfnamefont {S.}~\bibnamefont
  {Foffa}}\ and\ \bibinfo {author} {\bibfnamefont {R.}~\bibnamefont
  {Sturani}},\ }\href@noop {} {\bibfield  {journal} {\bibinfo  {journal} {in
  preparation}\ } (\bibinfo {year} {2021})}\BibitemShut {NoStop}%
\bibitem [{\citenamefont {Davydychev}\ and\ \citenamefont
  {Tausk}(1993)}]{Davydychev:1992mt}%
  \BibitemOpen
  \bibfield  {author} {\bibinfo {author} {\bibfnamefont {A.~I.}\ \bibnamefont
  {Davydychev}}\ and\ \bibinfo {author} {\bibfnamefont {J.~B.}\ \bibnamefont
  {Tausk}},\ }\href {\doibase 10.1016/0550-3213(93)90338-P} {\bibfield
  {journal} {\bibinfo  {journal} {Nucl. Phys.}\ }\textbf {\bibinfo {volume}
  {B397}},\ \bibinfo {pages} {123} (\bibinfo {year} {1993})}\BibitemShut
  {NoStop}%
\bibitem [{\citenamefont {Blanchet}(1995)}]{Blanchet:1995fr}%
  \BibitemOpen
  \bibfield  {author} {\bibinfo {author} {\bibfnamefont {L.}~\bibnamefont
  {Blanchet}},\ }\href {\doibase 10.1103/PhysRevD.51.2559} {\bibfield
  {journal} {\bibinfo  {journal} {Phys. Rev.}\ }\textbf {\bibinfo {volume}
  {D51}},\ \bibinfo {pages} {2559} (\bibinfo {year} {1995})},\ \Eprint
  {http://arxiv.org/abs/gr-qc/9501030} {arXiv:gr-qc/9501030 [gr-qc]}
  \BibitemShut {NoStop}%
\bibitem [{\citenamefont {Thorne}(1980)}]{Thorne:1980ru}%
  \BibitemOpen
  \bibfield  {author} {\bibinfo {author} {\bibfnamefont {K.~S.}\ \bibnamefont
  {Thorne}},\ }\href {\doibase 10.1103/RevModPhys.52.299} {\bibfield  {journal}
  {\bibinfo  {journal} {Rev. Mod. Phys.}\ }\textbf {\bibinfo {volume} {52}},\
  \bibinfo {pages} {299} (\bibinfo {year} {1980})}\BibitemShut {NoStop}%
\bibitem [{\citenamefont {Blanchet}\ \emph {et~al.}(2010)\citenamefont
  {Blanchet}, \citenamefont {Detweiler}, \citenamefont {Le~Tiec},\ and\
  \citenamefont {Whiting}}]{Blanchet:2010zd}%
  \BibitemOpen
  \bibfield  {author} {\bibinfo {author} {\bibfnamefont {L.}~\bibnamefont
  {Blanchet}}, \bibinfo {author} {\bibfnamefont {S.~L.}\ \bibnamefont
  {Detweiler}}, \bibinfo {author} {\bibfnamefont {A.}~\bibnamefont {Le~Tiec}},
  \ and\ \bibinfo {author} {\bibfnamefont {B.~F.}\ \bibnamefont {Whiting}},\
  }\href {\doibase 10.1103/PhysRevD.81.084033} {\bibfield  {journal} {\bibinfo
  {journal} {Phys. Rev.}\ }\textbf {\bibinfo {volume} {D81}},\ \bibinfo {pages}
  {084033} (\bibinfo {year} {2010})},\ \Eprint {http://arxiv.org/abs/1002.0726}
  {arXiv:1002.0726 [gr-qc]} \BibitemShut {NoStop}%
\bibitem [{\citenamefont {Goldberger}\ \emph {et~al.}(2014)\citenamefont
  {Goldberger}, \citenamefont {Ross},\ and\ \citenamefont
  {Rothstein}}]{Goldberger:2012kf}%
  \BibitemOpen
  \bibfield  {author} {\bibinfo {author} {\bibfnamefont {W.~D.}\ \bibnamefont
  {Goldberger}}, \bibinfo {author} {\bibfnamefont {A.}~\bibnamefont {Ross}}, \
  and\ \bibinfo {author} {\bibfnamefont {I.~Z.}\ \bibnamefont {Rothstein}},\
  }\href {\doibase 10.1103/PhysRevD.89.124033} {\bibfield  {journal} {\bibinfo
  {journal} {Phys. Rev.}\ }\textbf {\bibinfo {volume} {D89}},\ \bibinfo {pages}
  {124033} (\bibinfo {year} {2014})},\ \Eprint {http://arxiv.org/abs/1211.6095}
  {arXiv:1211.6095 [hep-th]} \BibitemShut {NoStop}%
\bibitem [{\citenamefont {Blanchet}(1998)}]{Blanchet:1997ji}%
  \BibitemOpen
  \bibfield  {author} {\bibinfo {author} {\bibfnamefont {L.}~\bibnamefont
  {Blanchet}},\ }\href {\doibase 10.1088/0264-9381/15/1/008} {\bibfield
  {journal} {\bibinfo  {journal} {Class. Quant. Grav.}\ }\textbf {\bibinfo
  {volume} {15}},\ \bibinfo {pages} {89} (\bibinfo {year} {1998})},\ \Eprint
  {http://arxiv.org/abs/gr-qc/9710037} {arXiv:gr-qc/9710037 [gr-qc]}
  \BibitemShut {NoStop}%
\bibitem [{\citenamefont {Punturo}\ \emph {et~al.}(2010)\citenamefont {Punturo}
  \emph {et~al.}}]{Punturo:2010zz}%
  \BibitemOpen
  \bibfield  {author} {\bibinfo {author} {\bibfnamefont {M.}~\bibnamefont
  {Punturo}} \emph {et~al.},\ }\bibfield  {booktitle} {\emph {\bibinfo
  {booktitle} {{Proceedings, 14th Workshop on Gravitational wave data analysis
  (GWDAW-14): Rome, Italy, January 26-29, 2010}}},\ }\href {\doibase
  10.1088/0264-9381/27/19/194002} {\bibfield  {journal} {\bibinfo  {journal}
  {Class. Quant. Grav.}\ }\textbf {\bibinfo {volume} {27}},\ \bibinfo {pages}
  {194002} (\bibinfo {year} {2010})}\BibitemShut {NoStop}%
\bibitem [{\citenamefont {Audley}\ \emph {et~al.}(2017)\citenamefont {Audley}
  \emph {et~al.}}]{Audley:2017drz}%
  \BibitemOpen
  \bibfield  {author} {\bibinfo {author} {\bibfnamefont {H.}~\bibnamefont
  {Audley}} \emph {et~al.} (\bibinfo {collaboration} {LISA}),\ }\href@noop {}
  {\  (\bibinfo {year} {2017})},\ \Eprint {http://arxiv.org/abs/1702.00786}
  {arXiv:1702.00786 [astro-ph.IM]} \BibitemShut {NoStop}%
\bibitem [{\citenamefont {Blanchet}\ \emph {et~al.}(2020)\citenamefont
  {Blanchet}, \citenamefont {Foffa}, \citenamefont {Larrouturou},\ and\
  \citenamefont {Sturani}}]{Blanchet:2019rjs}%
  \BibitemOpen
  \bibfield  {author} {\bibinfo {author} {\bibfnamefont {L.}~\bibnamefont
  {Blanchet}}, \bibinfo {author} {\bibfnamefont {S.}~\bibnamefont {Foffa}},
  \bibinfo {author} {\bibfnamefont {F.}~\bibnamefont {Larrouturou}}, \ and\
  \bibinfo {author} {\bibfnamefont {R.}~\bibnamefont {Sturani}},\ }\href
  {\doibase 10.1103/PhysRevD.101.084045} {\bibfield  {journal} {\bibinfo
  {journal} {Phys. Rev. D}\ }\textbf {\bibinfo {volume} {101}},\ \bibinfo
  {pages} {084045} (\bibinfo {year} {2020})},\ \Eprint
  {http://arxiv.org/abs/1912.12359} {arXiv:1912.12359 [gr-qc]} \BibitemShut
  {NoStop}%
\bibitem [{\citenamefont {Foffa}\ \emph
  {et~al.}(2019{\natexlab{b}})\citenamefont {Foffa}, \citenamefont {Mastrolia},
  \citenamefont {Sturani}, \citenamefont {Sturm},\ and\ \citenamefont
  {Torres~Bobadilla}}]{Foffa:2019hrb}%
  \BibitemOpen
  \bibfield  {author} {\bibinfo {author} {\bibfnamefont {S.}~\bibnamefont
  {Foffa}}, \bibinfo {author} {\bibfnamefont {P.}~\bibnamefont {Mastrolia}},
  \bibinfo {author} {\bibfnamefont {R.}~\bibnamefont {Sturani}}, \bibinfo
  {author} {\bibfnamefont {C.}~\bibnamefont {Sturm}}, \ and\ \bibinfo {author}
  {\bibfnamefont {W.~J.}\ \bibnamefont {Torres~Bobadilla}},\ }\href {\doibase
  10.1103/PhysRevLett.122.241605} {\bibfield  {journal} {\bibinfo  {journal}
  {Phys. Rev. Lett.}\ }\textbf {\bibinfo {volume} {122}},\ \bibinfo {pages}
  {241605} (\bibinfo {year} {2019}{\natexlab{b}})},\ \Eprint
  {http://arxiv.org/abs/1902.10571} {arXiv:1902.10571 [gr-qc]} \BibitemShut
  {NoStop}%
\bibitem [{\citenamefont {Bl{\"u}mlein}\ \emph {et~al.}(2020)\citenamefont
  {Bl{\"u}mlein}, \citenamefont {Maier},\ and\ \citenamefont
  {Marquard}}]{Blumlein:2019zku}%
  \BibitemOpen
  \bibfield  {author} {\bibinfo {author} {\bibfnamefont {J.}~\bibnamefont
  {Bl{\"u}mlein}}, \bibinfo {author} {\bibfnamefont {A.}~\bibnamefont {Maier}},
  \ and\ \bibinfo {author} {\bibfnamefont {P.}~\bibnamefont {Marquard}},\
  }\href {\doibase 10.1016/j.physletb.2019.135100} {\bibfield  {journal}
  {\bibinfo  {journal} {Phys. Lett.}\ }\textbf {\bibinfo {volume} {B800}},\
  \bibinfo {pages} {135100} (\bibinfo {year} {2020})},\ \Eprint
  {http://arxiv.org/abs/1902.11180} {arXiv:1902.11180 [gr-qc]} \BibitemShut
  {NoStop}%
\bibitem [{\citenamefont {Westpfahl}\ and\ \citenamefont
  {Goller}(1979)}]{Westpfahl:1979gu}%
  \BibitemOpen
  \bibfield  {author} {\bibinfo {author} {\bibfnamefont {K.}~\bibnamefont
  {Westpfahl}}\ and\ \bibinfo {author} {\bibfnamefont {M.}~\bibnamefont
  {Goller}},\ }\href {\doibase 10.1007/BF02817047} {\bibfield  {journal}
  {\bibinfo  {journal} {Lett. Nuovo Cim.}\ }\textbf {\bibinfo {volume} {26}},\
  \bibinfo {pages} {573} (\bibinfo {year} {1979})}\BibitemShut {NoStop}%
\bibitem [{\citenamefont {Ledvinka}\ \emph {et~al.}(2008)\citenamefont
  {Ledvinka}, \citenamefont {Sch{\"a}fer},\ and\ \citenamefont
  {Bicak}}]{Ledvinka:2008tk}%
  \BibitemOpen
  \bibfield  {author} {\bibinfo {author} {\bibfnamefont {T.}~\bibnamefont
  {Ledvinka}}, \bibinfo {author} {\bibfnamefont {G.}~\bibnamefont
  {Sch{\"a}fer}}, \ and\ \bibinfo {author} {\bibfnamefont {J.}~\bibnamefont
  {Bicak}},\ }\href {\doibase 10.1103/PhysRevLett.100.251101} {\bibfield
  {journal} {\bibinfo  {journal} {Phys. Rev. Lett.}\ }\textbf {\bibinfo
  {volume} {100}},\ \bibinfo {pages} {251101} (\bibinfo {year} {2008})},\
  \Eprint {http://arxiv.org/abs/0807.0214} {arXiv:0807.0214 [gr-qc]}
  \BibitemShut {NoStop}%
\bibitem [{\citenamefont {Foffa}(2014)}]{Foffa:2013gja}%
  \BibitemOpen
  \bibfield  {author} {\bibinfo {author} {\bibfnamefont {S.}~\bibnamefont
  {Foffa}},\ }\href {\doibase 10.1103/PhysRevD.89.024019} {\bibfield  {journal}
  {\bibinfo  {journal} {Phys. Rev.}\ }\textbf {\bibinfo {volume} {D89}},\
  \bibinfo {pages} {024019} (\bibinfo {year} {2014})},\ \Eprint
  {http://arxiv.org/abs/1309.3956} {arXiv:1309.3956 [gr-qc]} \BibitemShut
  {NoStop}%
\bibitem [{\citenamefont {Damour}(2016)}]{Damour:2016gwp}%
  \BibitemOpen
  \bibfield  {author} {\bibinfo {author} {\bibfnamefont {T.}~\bibnamefont
  {Damour}},\ }\href {\doibase 10.1103/PhysRevD.94.104015} {\bibfield
  {journal} {\bibinfo  {journal} {Phys. Rev.}\ }\textbf {\bibinfo {volume}
  {D94}},\ \bibinfo {pages} {104015} (\bibinfo {year} {2016})},\ \Eprint
  {http://arxiv.org/abs/1609.00354} {arXiv:1609.00354 [gr-qc]} \BibitemShut
  {NoStop}%
\bibitem [{\citenamefont {Blanchet}\ and\ \citenamefont
  {Fokas}(2018)}]{Blanchet:2018yvb}%
  \BibitemOpen
  \bibfield  {author} {\bibinfo {author} {\bibfnamefont {L.}~\bibnamefont
  {Blanchet}}\ and\ \bibinfo {author} {\bibfnamefont {A.~S.}\ \bibnamefont
  {Fokas}},\ }\href {\doibase 10.1103/PhysRevD.98.084005} {\bibfield  {journal}
  {\bibinfo  {journal} {Phys. Rev.}\ }\textbf {\bibinfo {volume} {D98}},\
  \bibinfo {pages} {084005} (\bibinfo {year} {2018})},\ \Eprint
  {http://arxiv.org/abs/1806.08347} {arXiv:1806.08347 [gr-qc]} \BibitemShut
  {NoStop}%
\bibitem [{\citenamefont {Damour}(2018)}]{Damour:2017zjx}%
  \BibitemOpen
  \bibfield  {author} {\bibinfo {author} {\bibfnamefont {T.}~\bibnamefont
  {Damour}},\ }\href {\doibase 10.1103/PhysRevD.97.044038} {\bibfield
  {journal} {\bibinfo  {journal} {Phys. Rev.}\ }\textbf {\bibinfo {volume}
  {D97}},\ \bibinfo {pages} {044038} (\bibinfo {year} {2018})},\ \Eprint
  {http://arxiv.org/abs/1710.10599} {arXiv:1710.10599 [gr-qc]} \BibitemShut
  {NoStop}%
\bibitem [{\citenamefont {Cheung}\ \emph {et~al.}(2018)\citenamefont {Cheung},
  \citenamefont {Rothstein},\ and\ \citenamefont {Solon}}]{Cheung:2018wkq}%
  \BibitemOpen
  \bibfield  {author} {\bibinfo {author} {\bibfnamefont {C.}~\bibnamefont
  {Cheung}}, \bibinfo {author} {\bibfnamefont {I.~Z.}\ \bibnamefont
  {Rothstein}}, \ and\ \bibinfo {author} {\bibfnamefont {M.~P.}\ \bibnamefont
  {Solon}},\ }\href {\doibase 10.1103/PhysRevLett.121.251101} {\bibfield
  {journal} {\bibinfo  {journal} {Phys. Rev. Lett.}\ }\textbf {\bibinfo
  {volume} {121}},\ \bibinfo {pages} {251101} (\bibinfo {year} {2018})},\
  \Eprint {http://arxiv.org/abs/1808.02489} {arXiv:1808.02489 [hep-th]}
  \BibitemShut {NoStop}%
\bibitem [{\citenamefont {Bern}\ \emph {et~al.}(2019)\citenamefont {Bern},
  \citenamefont {Cheung}, \citenamefont {Roiban}, \citenamefont {Shen},
  \citenamefont {Solon},\ and\ \citenamefont {Zeng}}]{Bern:2019nnu}%
  \BibitemOpen
  \bibfield  {author} {\bibinfo {author} {\bibfnamefont {Z.}~\bibnamefont
  {Bern}}, \bibinfo {author} {\bibfnamefont {C.}~\bibnamefont {Cheung}},
  \bibinfo {author} {\bibfnamefont {R.}~\bibnamefont {Roiban}}, \bibinfo
  {author} {\bibfnamefont {C.-H.}\ \bibnamefont {Shen}}, \bibinfo {author}
  {\bibfnamefont {M.~P.}\ \bibnamefont {Solon}}, \ and\ \bibinfo {author}
  {\bibfnamefont {M.}~\bibnamefont {Zeng}},\ }\href {\doibase
  10.1103/PhysRevLett.122.201603} {\bibfield  {journal} {\bibinfo  {journal}
  {Phys. Rev. Lett.}\ }\textbf {\bibinfo {volume} {122}},\ \bibinfo {pages}
  {201603} (\bibinfo {year} {2019})},\ \Eprint
  {http://arxiv.org/abs/1901.04424} {arXiv:1901.04424 [hep-th]} \BibitemShut
  {NoStop}%
\bibitem [{\citenamefont {Jantzen}(2011)}]{Jantzen:2011nz}%
  \BibitemOpen
  \bibfield  {author} {\bibinfo {author} {\bibfnamefont {B.}~\bibnamefont
  {Jantzen}},\ }\href {\doibase 10.1007/JHEP12(2011)076} {\bibfield  {journal}
  {\bibinfo  {journal} {JHEP}\ }\textbf {\bibinfo {volume} {12}},\ \bibinfo
  {pages} {076} (\bibinfo {year} {2011})},\ \Eprint
  {http://arxiv.org/abs/1111.2589} {arXiv:1111.2589 [hep-ph]} \BibitemShut
  {NoStop}%
\end{thebibliography}
%

\end{document}